 \documentclass[useAMS,usenatbib]{ashok}
 \usepackage{graphicx}
 \usepackage{epsfig}
\usepackage{amssymb}
\usepackage{amsmath}
\usepackage{aas_macros}

\title[{\bf{Geometry of halo and Lissajous orbits in the CRTBP}}]{Geometry of halo and Lissajous orbits in the circular restricted three-body problem 
with drag forces}
\author[{\bf{Ashok Kumar Pal and Badam Singh Kushvah}}]{Ashok Kumar Pal\thanks{E-mail:
ashokpalism@gmail.com (AKP)} and Badam Singh Kushvah\thanks{bskusk@gmail.com (BSK)}\\
Department of Applied Mathematics, Indian School of Mines, Dhanbad-826004, Jharkhand, India
}
\begin{document}

\date{}

\pagerange{\pageref{firstpage}--\pageref{lastpage}} \pubyear{2014}
\maketitle

\label{firstpage}
\begin{abstract}
In this paper, we determine the effect of radiation pressure, Poynting-Robertson drag and solar wind drag on the Sun-(Earth-Moon) 
restricted three body problem. Here, we take the larger body of the Sun as a larger primary, and Earth+Moon as a smaller primary. 
With the help of the perturbation technique, we find the Lagrangian points, and see that the collinear points deviate from 
the axis joining the primaries, whereas the triangular points remain unchanged in their configuration. We also find that 
 Lagrangian points move towards the Sun when radiation pressure increases. We have also analysed 
the stability of the triangular equilibrium points and have found that they are unstable because of the drag forces. Moreover, 
we have computed the halo orbits in the third-order approximation using Lindstedt-Poincar$\acute{e}$ method and have found the 
effect of the drag forces. According to this prevalence, the Sun-(Earth-Moon) model is used to design the trajectory for 
spacecraft traveling under the drag forces.
 \end{abstract} 
\section{Introduction}
During the last few years, many researchers have studied the effect of drag forces because of the significant role  they have 
in the dynamical system. For example, \cite{Murray1994Icar..112..465M} studied the dynamical 
effects of drag force in the circular restricted three body problem and found the approximate 
location and stability properties of the Lagrangian  points. \cite{Liou1995Icar..116..186L} examined the 
effects of radiation pressure, Poynting Robertson (P-R) drag and solar wind drag on dust grains trapped in 
mean motion resonance with the Sun-Jupiter restricted three-body problem. \cite{ishwar2006linear} studied the 
linear stability of triangular equilibrium points in the generalized photogravitational restricted three body problem 
with P-R drag and found that the triangular equilibrium points are unstable. Also, 
\cite{Kushvah2008Ap&SS.315..231K} determine the effect of radiation pressure on the equilibrium points in the generalized 
photogravitational restricted three body problem, and noticed that the collinear points deviate from the axis joining the 
two primaries, whereas the triangular points are not symmetric because of the presence of radiation pressure. Moreover, 
\cite{Kumari2013Ap&SS.344..347K} studied the motion of the infinitesimal mass in the restricted four body problem with solar 
wind drag and found the range of radiation factor of the equilibrium points.

We know that the Lagrangian points are important for mission design and transfer of trajectories. A number of missions 
have been successfully operated  in the vicinity of the Sun-Earth and Earth-Moon collinear Lagrangian points. In this 
regard, the {\it{International Sun-Earth Explorer (ISEE)}}, program was established as a joint project of the National 
Aeronautics and Space Administration (NASA) and the European Space Agency (ESA). {\it{ISEE-$3$}} was launched into a halo 
orbit around the Sun-Earth $L_1$ point in $1978$, allowing it to collect data on solar wind conditions upstream from the Earth. 
\cite{Farquhar1977JSpRo..14..170F} had designed the {\it{ISEE-$3$}} scientific satellite in the vicinity of the Sun-Earth 
interior Lagrangian point to continuously monitor the space between the Sun and the Earth.  {\it{WIND}} was launched on $1994$ 
November $1$ and was positioned in a sunward orbit. The {\it{Solar and Heliospheric Observatory (SOHO)}} project was launched 
in $1995$ December to study the internal structure of the Sun. The {\it{Advanced Composition Explorer (ACE)}} was launched 
in $1997$ and orbits the $L_1$ Lagrangian point, which is a point of the Sun-Earth gravitational equilibrium about $1.5$ 
million km from the Earth and $148.5$ million km from the Sun. {\it{ARTHEMIS}} was the first spacecraftto be in the vicinity 
of Earth-Moon Lagrangian point. In $2010$ August, the {\it{ARTHEMIS P$1$}} spacecraft entered an orbit near the 
Earth-Moon $L_{2}$ point for approximately $131$ d, before transferring to an $L_{1}$ quasi-halo orbit where it remained for an 
additional $85$ d. On $2011$ July $17$, the {\it{ARTHEMIS P$2$}} spacecraft was successfully inserted into the Earth-Moon 
Lagrangian point orbit with an arrival near the $L_{2}$ point in $2010$  October \citep{pavlak2012strategy}. The 
{\it{ARTHEMIS}} Lagrangian point orbit design features quasi-halo orbits demonstrated recently by \cite{folta2013preliminary}. 

To our knowledge, there have been numerous published papers that have extensively covered topics related to halo orbits, 
Lagrangian point satellite operations in general, and their applications within the Earth-Moon and the Sun-Earth system. 
For more information on halo orbits, we refer to the three-dimensional periodic halo orbits near the collinear Lagrangian points
in the restricted three-body problem obtained by \cite{Howell1984CeMec..32...53H}. She obtained orbits that increase in size 
when increasing the mass parameter $\mu$. \cite{Clarke2003AAS...203.0305C,Clarke2005APS..APR.C9007C} discussed a discovery 
mission concept that utilize occultations from a lunar halo orbit by the Moon to enable detection of terrestrial planets. 
\cite{Breakwell1979..20..389L} has computed halo orbits around the Earth-Moon $L_{2}$ point. \cite{Calleja2012CeMDA.114...77C} 
computed the unstable manifolds of selected vertical and halo orbits, which in several cases have led to the detection of 
heteroclinic connections from such a periodic orbit to invariant tori. Other authors have carried out similar work 
\citep{di1992nonlinear,farquhar2001flight,kim2001lyapunov,junge2002identification,kolemen2007quasi,hill2008autonomous}. However, 
they ignored the drag force, although, \cite{Eapen2014Ap&SS.352..437E} have studied the halo orbits at the 
Sun-Mars $L_1$ Lagrangian point in the photogravitational restricted three-body problem and have found that as the 
radiation pressure increases, the transition from Mars-centric path to heliocentric path is delayed. 

In this paper, we study the effect of radiation pressure, P-R drag, and solar wind drag  on the Lagrangian  
points and use the Lindstedt-Poincar$\acute{e}$ method to compute halo orbits in vicinity of the $L_1$ point of the 
Sun-Earth-Moon system. This paper is organized as follows. In Section $2$, we recall some well-known facts about the 
circular restricted three-body problem with drag forces (i.e. its equations of motion, equilibrium points,
and stability). In Section $3$, we describe  the motion near the Lagrangian point $L_{1}$, and use the 
Lindstedt-Poincar$\acute{e}$ method to compute the halo orbits. In Section $4$, we discuss our results. Finally, in Appendix A, 
we provide all the coefficients.
\section{Mathematical formulation of the problem and equations of motion}
We formulate the Sun-Earth-Moon system with the radiation pressure, P-R drag 
and solar wind drag \citep{Parker1965SSRv....4..666P} and proceed with the problem as follows \citep{Liou1995Icar..116..186L}. 
The equations of motion in the rotating reference frame are
  \begin{eqnarray}
  && \ddot{x}-2 \dot{y}=\frac{\partial{U}}{\partial{x}}+(1+sw)F_x, \label{eq:1}\\&&
   \ddot{y}+2\dot{x}=\frac{\partial{U}}{\partial{y}}+(1+sw)F_y, \label{eq:2}\\&&
   \ddot{z}=\frac{\partial{U}}{\partial{z}}+(1+sw)F_z \label{eq:3},
  \end{eqnarray}
where 
\begin{eqnarray}
 &&U=\frac{(1-\beta)(1-\mu)}{r_1}+\frac{\mu}{r_2}+\frac{1}{2}(x^2 +y^2),
\end{eqnarray}
and drag force components are 
\begin{eqnarray}
&& F_x=\frac{-\beta (1-\mu)}{cr_{1}^2}\left\{\frac{[(x+\mu)\dot{x}+(\dot{y}-\mu)y+z\dot{z}](x+\mu)}{r_{1}^2}
 \right. \nonumber\\&&  \left.+ (\dot{x}-y)\right\}, \\&&
 F_y=\frac{-\beta (1-\mu)}{cr_{1}^2}\left\{\frac{[(x+\mu)\dot{x}+(\dot{y}-\mu)y+z\dot{z}]y}{r_{1}^2}+
 \right.\nonumber\\&& \left.+ (\dot{y}+x)\right\},\\&&
 F_{z}=\frac{-\beta (1-\mu)}{cr_{1}^2}\left\{\frac{[(x+\mu)\dot{x}+(\dot{y}-\mu)y+z\dot{z}]z}{r_{1}^2}+\dot{z}\right\}.\nonumber\\&&
\end{eqnarray}
It is supposed that $m_1$ is the mass of the Sun, and $m_2$ is the mass of the Earth plus Moon, and hence the mass parameter  
 $\mu=m_{2}/(m_{1}+m_{2})$. However, the definition of mass parameter $\mu$ is different from that of 
 \citep{Liou1995Icar..116..186L}. They have used $\mu_{1}$ and $\mu_{2}$ to represent the masses of the Sun and Jupiter, 
 respectively, whereas in the problem the masses are represented by $m_{1}$ and $m_{2}$. Here, $\beta$ is the ratio of 
 radiation pressure force to the solar gravitation force, $sw$ is the ratio of solar wind drag to P-R drag 
 \citep{Burns1979Icar...40....1B, Gustafson1994AREPS..22..553G} and $c$ is the unitless speed of light. The unit of mass is 
 taken in such a way that $G(m_1+m_2)=1$; the unit of distance is taken as  the distance of the center of mass 
 (of the Earth-Moon) to the Sun, whereas the unit of time is taken to be the time period of the rotating frame. 

In order to calculate the Lagrangian equilibrium points, we solve equations \eqref{eq:1}-\eqref{eq:3} with the condition 
that all derivatives are zero, and we obtain
\begin{eqnarray}
 &&x-\frac{(1-\beta)(1-\mu)}{r_{1}^3}(x+\mu)-\frac{\mu}{r_{2}^3}(x+\mu-1)\nonumber\\&&+
 \frac{(1+sw)\beta(1-\mu)}{r_{1}^2}\left[\frac{\mu y(x+\mu)}{cr_{1}^2}+\frac{y}{c}\right]=0,\label{eq:4}\\&&
 y\left[1-\frac{(1-\beta)(1-\mu)}{r_{1}^3}-\frac{\mu}{r_{2}^3}\right]+\frac{(1+sw)\beta(1-\mu)}{r_{1}^2}\nonumber\\&&\times
 \left[\frac{\mu y^2}{c r_{1}^2}-\frac{x}{c}\right]=0,\label{eq:5}\\&&
 z\left[\frac{(1-\beta)(1-\mu)}{r_{1}^3}+\frac{\mu}{r_{2}^3}- \frac{(1+sw)\beta(1-\mu)\mu y}{cr_{1}^4}\right]=0.\nonumber\\\label{eq:6}
\end{eqnarray}
From equation (\ref{eq:6}), we obtain two possible solutions, either $z=0$ or $z\ne0$. If $z\ne0$ then 
\begin{eqnarray}
 &&\frac{(1-\beta)(1-\mu)}{r_{1}^3}+\frac{\mu}{r_{2}^3}= \frac{(1+sw)\beta(1-\mu)\mu y}{c r_{1}^4}.\label{eq:coz}
\end{eqnarray}
From equation (\ref{eq:coz}), with $\beta=0$, (i.e., no radiation pressure is taken into account), we obtain
\begin{eqnarray}
 &&\frac{1-\mu}{r_{1}^3}=-\frac{\mu}{r_{2}^3}.\label{eq:coz1}
\end{eqnarray}
From the right-hand side of equation (\ref{eq:coz1}), because $\mu>0$, and also $r_{2}$ is the distance of the infinitesimal 
body from the second primary, we find that
\begin{eqnarray}
 &&\frac{1-\mu}{r_{1}^{3}}=-\frac{\mu}{r_{2}^3}<0.
\end{eqnarray}
This gives $\mu>1$, which is never possible. Therefore, at $\beta=0$, there are no equilibrium points outside the $xy$-plane.

Again, with $z\ne0$ and if $0<\beta<1$, then it is obvious from equation \eqref{eq:coz} that 
 the only possible solution is to have $y>0$. From equation \eqref{eq:5} and \eqref{eq:coz}, we obtain 
  \begin{eqnarray}
   &&y-\frac{(1+sw)\beta(1-\mu)x}{cr_{1}^2}=0.\label{eq:a1}
  \end{eqnarray}
 With the condition that $y>0$, equation \eqref{eq:a1} gives one possible solution $x>0$, and therefore $x+\mu>0$.
 Now, we divide both sides of equation \eqref{eq:4} by $(x+\mu)$, and using equation \eqref{eq:coz}, we obtain
  \begin{eqnarray}
   \frac{x}{x+\mu}+\frac{\mu}{r_{2}^3 (x+\mu)}+\frac{(1+sw)\beta(1-\mu)y}{c r_{1}^2(x+\mu)}=0.\label{eq:az}
  \end{eqnarray}
  For $x>0$ and $y>0$, the left-hand side of equation \eqref{eq:az} has a non zero quantity. Therefore, $z$ must 
  be zero and have no equilibrium points outside the $xy$-plane.

 \cite{Simmons1985CeMec..35..145S} have shown that out-of-plane equilibrium points occur if 
 $(1-\beta_{1})/(1-\beta_{2})<0$, where $\beta_{1,2}$ are the ratio of the magnitudes of radiation to the gravitational 
 forces from $m_{1,2}$. However, in the present Sun-Earth-Moon system, only the Sun is a radiating body. Therefore, 
 $\beta_{2}=0$ and $\beta_{1}=\beta >1$, which is not possible. Consequently, out-of-plane equilibrium points do not exist 
 in the present model. 
 
Thus, for $z=0$, we solve equations (\ref{eq:4}) and (\ref{eq:5}) using Taylor series expansions in $r_{1}$ and 
$r_{2}$ and some approximations, as in \citep{murray1999solar}. The general solutions are given by
\begin{eqnarray}
&& x^{*}=x_{0}+\Delta x,  \ \mbox{and} \ y^{*}=y_{0}+\Delta y. \label{eq:x}
\end{eqnarray}
Here, $\Delta x$ and $\Delta y$ are the small quantities that are introduced for drag forces and 
$(x_{0}, y_{0})$ is a solution of the equations when there is no drag force. The corresponding distances of the infinitesimal 
body to the masses $m_{1}$ and $m_{2}$ are given by 
\begin{eqnarray}
&& r_{1}^{*}=\sqrt{(x^{*}+\mu)^2 +y^{*2}+z^{*2}}, \\  \mbox{and} \  &&r_{2}^{*}=\sqrt{(x^{*}+\mu-1)^2 +y^{*2}+z^{*2}}.
\end{eqnarray}
We use the Taylor series expansions around $(x_{0}, y_{0})$ and neglect second- and 
higher-order terms in $\Delta x$ and $\Delta y$. Then, we solve the simultaneous equations in $\Delta x$ 
and $\Delta y$ ($\because  z_{0}=0$ for all Lagrangian points). We obtain following expressions of $\Delta x$
 and $\Delta y$ for all the Lagrangian points with a fixed value of $sw=0.35$ \citep{Gustafson1994AREPS..22..553G}:\\
for $L_{1}$,
\begin{eqnarray}
 &&\Delta x=0.494997-\frac{2.19061}{4.42551-\beta},\nonumber\\&&
 \Delta y=\frac{0.470726(-3.74923\times10^{-9}+4.23594\times10^{-10}\beta)}{(-4.42551+\beta)(-2.97008+\beta)},\nonumber
\end{eqnarray}
for $L_{2}$,
\begin{eqnarray}
 &&\Delta x=\frac{0.505037(-1.23265\times10^{-15}+\beta)}{-4.57611+\beta},\nonumber\\&&
 \Delta y=\frac{0.530994(-3.57762\times10^{-9}+3.90903\times10^{-10}\beta)\beta}{(-4.57611+\beta)(-3.03032+\beta)},\nonumber
\end{eqnarray} 
for $L_{3}$,
\begin{eqnarray}
 &&\Delta x=\frac{0.5 \beta}{1.5-\beta},\nonumber\\&&
 \Delta y=\frac{0.500004(1.22068\times10^{-9}-4.06893\times10^{-10})\beta}{(-1.5+\beta)(3.34357\times10^{-6}+\beta)},\nonumber
\end{eqnarray}
for $L_{4,5}$,
\begin{eqnarray}
 &&\Delta x=\frac{0.25\beta(0.0000126643+\beta)(3.00001+\beta)}{(-1.5+\beta)(5.83558\times10^{-6}+\beta)(3.00001+\beta)},
 \nonumber\\&& \label{eq:lx}\\
 &&\Delta y=\pm\frac{0.500002(3.08248\times10^{-6}+0.866021\beta)\beta}{(-1.5+\beta)(5.83558\times10^{-6}+\beta)}.\label{eq:ly}
\end{eqnarray}
 
  \begin{figure}
\begin{center} 
   \includegraphics[width=8cm]{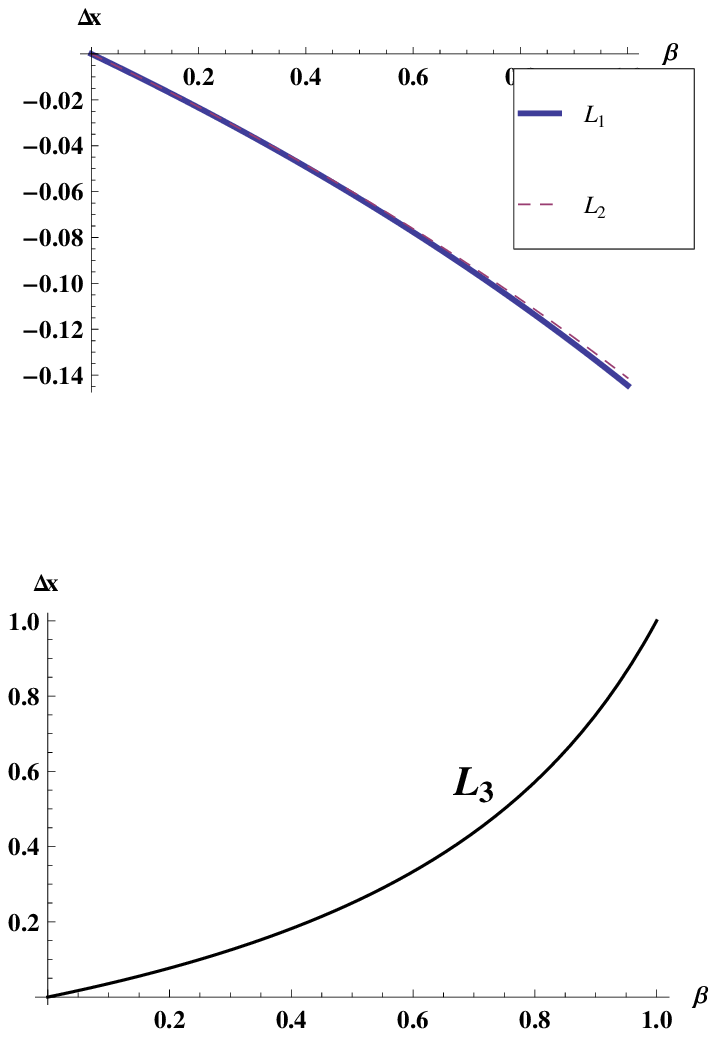}
  \caption{The effect of $\beta$ on $x$ coordinate in $L_{1,2,3}$ points} 
  \label{fig:colx}
  \end{center}
 \end{figure}
 \begin{figure} 
 \begin{center} 
 \includegraphics[width=8cm]{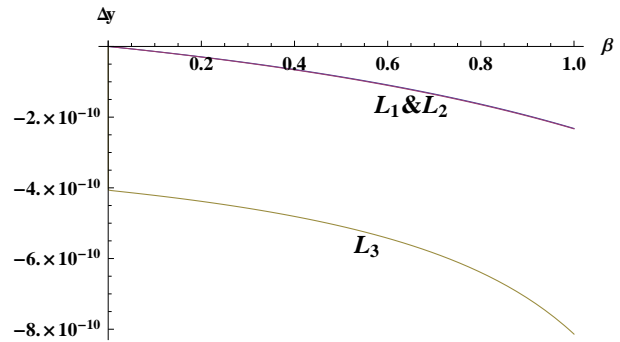}
   \caption{The effect of $\beta$ on $y$ coordinate in $L_{1,2,3}$ points} 
   \label{fig:coly}
   \end{center}
  \end{figure}
  The effect of $\beta$ in $\Delta x$ on the $L_{1}$ and $L_{2}$ points is shown in the top panel of Fig. \ref{fig:colx},
  whereas the bottom panel is for the $L_{3}$ point. For the $L_{1}$ and $L_{2}$ points, the effect of $\beta$ is approximately 
  equal and $\Delta x$ is increasing negatively (i.e. both the $L_{1}$ and $L_{2}$ points tend towards the Sun with increasing 
  radiation pressure). For the $L_{3}$ point, $\Delta x$ increases positively when $\beta$ increases, and therefore the 
  Lagrangian point $L_{3}$ also tends towards the Sun. $\Delta y$ corresponds to the case when $L_{1,2,3}$ increases negatively, 
  for an increasing value of $\beta$, which is shown in Fig. \ref{fig:coly}. Therefore, when the radiation pressure increases, 
  the collinear points perturb from their collinearity and tend towards the radiating body of the Sun.  
  
 \begin{table*}
  \centering
  \begin{minipage}{140mm}
   \caption{Lagrangian equilibrium points $L_{i},(i=1,2,3)$ with $s_w=0.35$.}\label{tab:1}
   \begin{tabular}{@{}cccccc@{}}
   \hline
$\beta$&$L_{1}$&$L_{2}$&$L_{3}$\\
\hline
 0.0 & $(0.989991,~0)$ & $(1.01007,~0)$ & $(-1.00000326495,0)$ \\
 0.1 & $(0.978547,~-1.40554\times 10^{-11})$ & $(0.998787,~-1.43251\times 10^{-11})$ & $(-0.964289,~-4.21415\times 10^{-10})$\\
 0.2 & $(0.966562,~-2.94742\times 10^{-11})$ & $(0.986989,~-3.0005\times 10^{-11})$ & $(-0.92308,~-4.3819\times10^{-10})$\\
 0.3 & $(0.953996,~-4.64359\times10^{-11})$ & $(0.974638,~-4.72137\times10^{-11})$ & $(-0.875003,~-4.57754\times10^{-10})$\\
 0.4 & $(0.940805,~-6.51505\times10^{-11})$ & $(0.961696,~-6.6153\times10^{-11})$ & $(-0.818185,~-4.80874\times10^{-10})$\\
 0.5 & $(0.926942,~-8.58655\times10^{-11})$ & $(0.948119,~-8.70629\times10^{-11})$ & $(-0.750004,~-5.08618\times10^{-10})$ \\
 0.6 & $(0.912355,~-1.08874\times10^{-10})$ & $(0.93386,~-1.10221\times10^{-10})$ & $(-0.666671,~-5.42526\times10^{-10})$\\
 0.7 & $(0.896984,~-1.34524\times10^{-10})$ & $(0.918864,~-1.35961\times10^{-10})$ & $(-0.562504,~-5.84912\times10^{-10})$\\
 0.8 & $(0.880766,~-1.63235\times10^{-10})$ & $(0.903074,~-1.64679\times10^{-10})$ & $(-0.428576,-6.39407\times10^{-10})$\\
 0.9 & $(0.863627,~-1.95512\times10^{-10})$ & $(0.886425,~-1.96851\times10^{-10})$ & $(-0.250006,~-7.12067\times10^{-10})$\\
 1.0 & $(0.845488,~-2.31971\times10^{-10})$ & $(0.868845,~-2.33054\times10^{-10})$ & $(-7.51792\times10^{-6},~-8.13791\times10^{-10})$\\
\hline
 \end{tabular}
\end{minipage}
 \end{table*}   
  The different Lagrangian points $L_{i}, \ (i=1,2,3)$ at different values of $\beta$ are presented in Table \ref{tab:1}. 
 \begin{figure}
\begin{center} 
   \includegraphics[width=6cm]{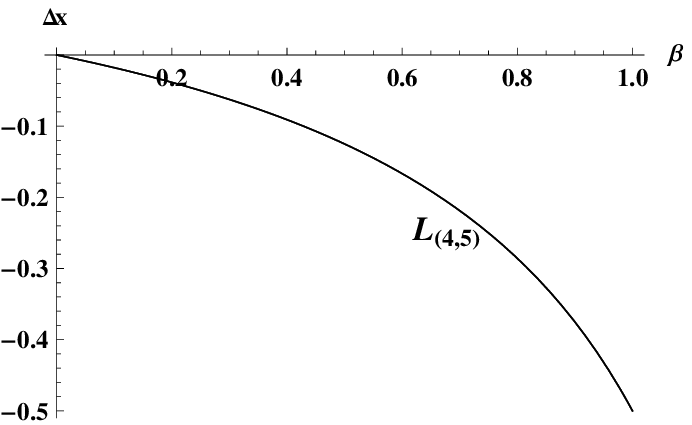}
  \caption{The effect of $\beta$ on $x$ coordinate of triangular points} 
  \label{fig:trix}
  \end{center}
 \end{figure}
 \begin{figure} 
 \begin{center} 
 \includegraphics[width=6cm]{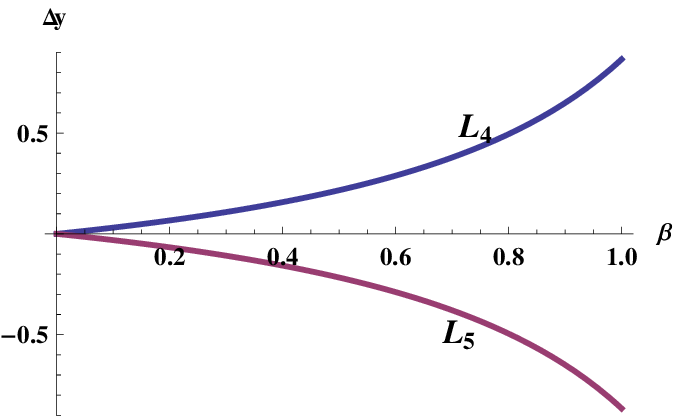}
   \caption{The effect of $\beta$ on $y$ coordinate of triangular points} 
   \label{fig:triy}
   \end{center}
  \end{figure}
The effect of radiation pressure in $\Delta x$ of $L_{4,5}$ is the same, as shown in Fig. \ref{fig:trix}, while 
Fig. \ref{fig:triy} shows that there is a symmetrical change in $\Delta y$ of $L_{4}$ and $L_{5}$. These equilibrium points 
also tend towards the Sun symmetrically when $\beta$ increases.  
\begin{table*}
  \centering
  \begin{minipage}{140mm}
   \caption{Triangular equilibrium points with $sw=0.35$}\label{tbl:2}
   \begin{tabular}{@{}ccccc@{}}
   \hline
$\beta$&$L_{(4,~5)}$\\
\hline
0.0 & $(0.499997,~\pm0.866025)$\\
0.1 & $(0.482139,~\pm0.835096)$\\
0.2 & $(0.461534,~\pm0.799408)$\\
0.3 & $(0.437496,~\pm0.757773)$\\
0.4 & $(0.409086,~\pm0.708567)$\\
0.5 & $(0.374995,~\pm0.64952)$ \\
0.6 & $(0.333329,~\pm0.577351)$\\
0.7 & $(0.281245,~\pm0.487141)$\\
0.8 & $(0.214281,~\pm0.371155)$\\
0.9 & $(0.124995,~\pm0.216509)$\\
1.0 & $(-5.81166\times10^{-6},~\pm3.08134\times10^{-6})$\\
\hline
 \end{tabular}
\end{minipage}
 \end{table*} 
 The different triangular points $L_{4,5}$ at different values of $\beta$ are shown in Table \ref{tbl:2}.
 
\subsection{Stability of the equilibrium points}
The location of equilibrium points do not affect our knowledge of their stability. We now consider the stability  
property by using the standard technique of linearizing the perturbation equations in the vicinity of an equilibrium point. 
Our approach is based on that of \cite{Schuerman1980ApJ...238..337S} and \cite{Murray1994Icar..112..465M}.

Consider a small displacement from the equilibrium position ($x^*,y^*$), and let the solution for the subsequent motion be 
of the form $x=x^*+X$, \ $y=y^*+Y$, where
\begin{eqnarray}
 &&X=X_{0}e^{\lambda t}, \ \ Y=Y_{0}e^{\lambda t},
\end{eqnarray}
and $X_{0}$, $Y_{0}$ and $\lambda$, are constants. Using these substitutions in equations (\ref{eq:4}) 
and  (\ref{eq:5}), and using Taylor series expansion, the following simultaneous linear equations in $X$ and 
$Y$ are given as
\begin{eqnarray}
&& X \left[ \lambda^{2} +\frac{(1-\beta)(1-\mu)}{r_{1}^{*3}}\left(1-\frac{3(x^*+\mu)^2}{r_{1}^{*2}}\right)+
 \frac{\mu}{r_{2}^{*3}}\right.\nonumber\\&&\left.\times\left(1-\frac{3(x^*+\mu-1)^2}{r_{2}^{*3}}\right)-1-(1+sw)
 \left\{\left(\lambda\frac{\partial F_{x}}{\partial \dot{x}}\right)_{*}\right.\right.\nonumber\\&&\left.\left.+
 \left(\frac{\partial F_{x}}{\partial x}\right)_{*}\right\}\right]+Y\left[-2\lambda-
 \frac{3(1-\beta)(1-\mu)y^*(x^*+\mu)}{r_{1}^{*5}} \right.\nonumber\\&&\left. -\frac{3\mu y^*(x^*+\mu-1)}
 {r_{2}^{*5}}-(1+sw)\left\{\left(\lambda \frac{\partial F_{x}}{\partial \dot{y}}\right)_{*}\right.\right.\nonumber\\&&\left.\left. 
 +\left(\frac{\partial F_{y}}{\partial y}\right)_{*}\right\}\right]=0\label{eq:ri}
\end{eqnarray}
and
\begin{eqnarray}
 &&X\left[2\lambda-\frac{3(1-\beta)(1-\mu)y^{*}(x^*+\mu)}{r_{1}^{*5}}-\frac{3\mu y^*(x+\mu-1)}{r_{2}^{*5}}\right.\nonumber
 \\&&\left.-(1+sw)\left\{\left(\lambda\frac{\partial F_{y}}{\partial \dot{x}}\right)_{*}-\left(
 \frac{\partial F_{y}}{\partial x}\right)\right\}\right]+Y\left[\lambda^2\right.\nonumber\\&&\left.+\frac{(1-\mu)(1-\beta)}
 {r_{1}^{*3}}\left(1-\frac{3y^{*2}}{r_{1}^{*2}}\right)+\frac{\mu}{r_{2}^{*3}}\left(1-\frac{3y^{*2}}{r_{2}^{*2}}\right)
 \right.\nonumber\\&&\left.-1-(1+sw)\left\{\lambda\left(\frac{\partial F_{y}}{\partial \dot{y}}\right)_{*}
 -\left(\frac{\partial F_{y}}{\partial y}\right)_{*}\right\}\right]=0,\label{eq:13}
\end{eqnarray}
where $\left(\right)_{*}$ denotes the evaluation of a partial derivative at the displaced equilibrium point. We can now 
rewrite equations (\ref{eq:ri}) and (\ref{eq:13}) as
\begin{eqnarray}
 &&X[\lambda^2+a^*-d^*-1-(1+sw)(\lambda K_{x,\dot{x}}+K_{x,x})]\nonumber\\&&+Y[-2\lambda-c^*-(1+sw)
 (\lambda K_{x,\dot{y}}+K_{x,y})]=0,\label{eq:14}
 \end{eqnarray}
 \begin{eqnarray}
 && X[2\lambda-c^*-(1+sw)(\lambda K_{y,\dot{x}}+K_{y,x})]+Y[\lambda^2 \nonumber\\&& +a^*-b^*-1-(1+sw)(\lambda K_{y,\dot{y}}
 +K_{y,y})]=0,\label{eq:15}
\end{eqnarray}
where the constants are
\begin{eqnarray}
 &&a^*=\frac{(1-\beta)(1-\mu)}{r_{1}^{*3}}+\frac{\mu}{r_{2}^{*3}},\\&&
 b^*=3\left[\frac{(1-\beta)(1-\mu)}{r_{1}^{*5}}+\frac{\mu}{r_{2}^{*5}}\right]y^{*2},\\&&
 c^*=3\left[\frac{(1-\beta)(1-\mu)(x^*+\mu)}{r_{1}^{*5}}+\frac{\mu(x^*+\mu-1)}{r_{2}^{*5}}
 \right]y^*,\nonumber\\&&\\&&
 d^*=3\left[\frac{(1-\beta)(1-\mu)(x^*+\mu)^2}{r_{1}^{*5}}+\frac{\mu(x^*+\mu-1)^2}{r_{2}^{*5}}
 \right],\nonumber\\&&
\end{eqnarray}
and
\begin{eqnarray}
&& K_{x,x}=\left(\frac{\partial F_{x}}{\partial x}\right)_{*}, \   K_{x,\dot{x}}=\left(\frac{\partial F_{x}}{\partial 
 \dot{x}}\right)_{*},\label{eq:k1} \\ && K_{x,y}=\left(\frac{\partial F_{x}}{\partial y}\right)_{*}, \  K_{x,\dot{y}}=\left(\frac{\partial 
 F_{x}}{\partial \dot{y}}\right)_{*},\\&&  K_{y,x}=\left(\frac{\partial F_{y}}{\partial x}\right)_{*}, \ 
  K_{y,\dot{x}}=\left(\frac{\partial F_{y}}{\partial \dot{x}}\right)_{*},\\&&K_{y,y}=\left(\frac{\partial F_{y}}{\partial y}
   \right)_{*}, \  K_{y,\dot{y}}=\left(\frac{\partial F_{y}}{\partial \dot{y}}\right)_{*}\label{eq:k2}.
\end{eqnarray}
The condition for determinant of the linear equations defined by equations (\ref{eq:14}) and (\ref{eq:15}) needs to be 
zero. Neglecting the terms of $O(K^2)$, we obtain the characteristic equation as
\begin{eqnarray}
 \lambda^4+a_{3}\lambda^3+(a_{20}+a_2)\lambda^2+a_1\lambda+(a_{00}+a_0)=0,\label{eq:17}
\end{eqnarray}
where the approximate expressions for constants $a_{00}$ and $a_{20}$, and the drag force terms $a_{i}(i=0,1,2,3)$ are given 
in Appendix A.

If we take $K=0$, then the characteristic equation reduces to the following form 
\begin{eqnarray}
 &&\lambda^4+a_{20}\lambda^2+a_{00}=0,
\end{eqnarray}
which gives the classical solutions. The four roots in the classical problem occur as real and pure imaginary 
pair, as in the case of each $L_{i}, \ \  (i=1, 2, 3 )$, whereas two pure imaginary pairs occur in the case of $L_4$ and $L_{5}$. 
The stability of the given equilibrium point depends on the sign and value of the roots of the characteristic equation. If any 
of the roots has a positive real part, the motion is unstable to small displacements with exponential 
growth, whereas in classical case, the $L_{4}$ and $L_{5}$ points are linearly stable to small displacements in the system.

If we restrict our analysis to $L_{4}$ and $L_{5}$, the four roots of the characteristic equation, without drag forces,  
are given as  
\begin{eqnarray}
 &&\lambda_{n}=\pm Z i, \  \ (n=1,2,3,4)
\end{eqnarray}
where 
\begin{eqnarray}
&& Z=\sqrt{\frac{a_{20}\pm\sqrt{a_{20}^2-4a_{00}}}{2}}.
\end{eqnarray}
 In the presence of drag forces, we assume that the roots are in the following form 
\begin{eqnarray}
 &&\lambda=\eta Z\pm i(1+\gamma)Z,
\end{eqnarray}
that is, 
\begin{eqnarray}
&&\lambda_{1,2}=\eta Z+(1+\gamma)Z i\label{eq:ri1}, \\&& \lambda_{3,4}=\eta Z-(1+\gamma)Z i\label{eq:ri3},
\end{eqnarray}
where $\gamma$ and $\eta$ are small real quantities. With the help of equations (\ref{eq:17}), (\ref{eq:ri1}) and (\ref{eq:ri3}),  
neglecting the product of $\gamma$ and then $\eta$ with $a_{i}$, and solving the real and imaginary parts, we obtain
\begin{eqnarray}
  &&\eta=\frac{(a_{3}Z^2- a_1)}{2Z(-2Z^2+a_{20})},
\end{eqnarray}
and 
\begin{eqnarray}
  &&\gamma=\frac{(a_{00}+a_{0})-(a_{20}+a_{2})Z^2+Z^4}{2Z^2(a_{20}-2Z^2)}.
\end{eqnarray}
In all the cases, the real part of at least one characteristic root is positive. Therefore, the equilibrium point is saddle point.
\section{Computation of halo orbit} 
In order to discuss the motion near the Lagrangian point of the system, we choose a coordinate 
system centred at the Lagrangian point $L_{1}$ in the rotating reference frame. The equations of motion 
of an infinitesimal body are obtained by translating the origin to the location of $L_{1}$. The referred translation is given as
\begin{eqnarray}
 &&x=X-1+\mu+\gamma ,\   \ y=Y-l , \ \mbox{and} \ z=Z.
\end{eqnarray}
In this new coordinate system, the variables $x$, $y$ and $z$ are scaled. The distances between $L_1$ and the smaller primary 
are $\gamma$ in the $x$-axis and $l$ in the $y$-axis. Using these facts in equations (\ref{eq:1})-(\ref{eq:3}), 
we obtain the equations of motion with the help of Legendre polynomials for expanding the non-linear terms and considering 
the linear parts, and we can write
\begin{eqnarray}
 &&\ddot{x}-2\dot{y}-x \nu_{10}+y \nu_{11}=0,\label{eq:h1}\\&&
 \ddot{y}+2 \dot{x}-y\nu_{21}-x \nu_{20}=0,\label{eq:h2}\\&&
 \ddot{z}+\nu_{*}z=0.
\end{eqnarray}
All the coefficients are given in Appendix A.  
Here, the $z$-axis solution is simple harmonic, because $\nu_{*}>0$. However, the motion in the $xy$-plane is coupled. Solving 
equations (\ref{eq:h1}) and (\ref{eq:h2}), the characteristic equation has two real and two complex roots. The complex 
roots are not pure imaginary but the real parts of the complex roots are very small with respect to the age of solar system. 
Therefore, we neglect them and the roots are  $\pm\alpha$ and $\pm i\lambda$, where
\begin{eqnarray}
\alpha=\pm\sqrt{\frac{-4+\nu_{21}+\nu_{10}+\sqrt{(4-\nu_{21}-\nu_{10})^2 -4\zeta_{1}}}{2}},
\end{eqnarray}
and 
\begin{eqnarray}
 \lambda=\pm \sqrt{\frac{4-(\nu_{21}+\nu_{10})+\sqrt{(4-\nu_{21}-\nu_{10})^2-4\zeta_{1}}}{2}}.
\end{eqnarray}
 All the coefficients are given in Appendix A. 
 Because the two real roots are opposite in sign, arbitrarily chosen initial conditions give rise, to 
unbounded solutions as time increases. If, however, the initial conditions are restricted and only a non-divergent mode 
is allowed, the $xy$-solution will be bounded. In this case, the linearized equations have solutions of the form
\begin{eqnarray}
 &&x(t)=-A_{x}\cos(\lambda t+\phi),
  \end{eqnarray}
  \begin{eqnarray}
 && y(t)=\kappa A_{x}\{2\lambda \sin(\lambda t+\phi)-\nu_{11}\cos(\lambda t+
 \phi)\},
  \end{eqnarray}
  \begin{eqnarray}
 && z(t)=A_{z}\sin(\sqrt{\nu_{*}}t+\psi),
\end{eqnarray}
with 
\begin{eqnarray}
 &&\kappa=\frac{\lambda ^2 +\nu_{10}}{4\lambda^2 +\nu_{11}^2}.
\end{eqnarray}
The in-plane and out-of-plane frequencies are not equal, they are incommensurable. Then the linearized motion produces the 
Lissajous-type trajectories for the Sun-(Earth-Moon) system around $L_{1}$.
  \begin{figure}
 \includegraphics[width=7cm]{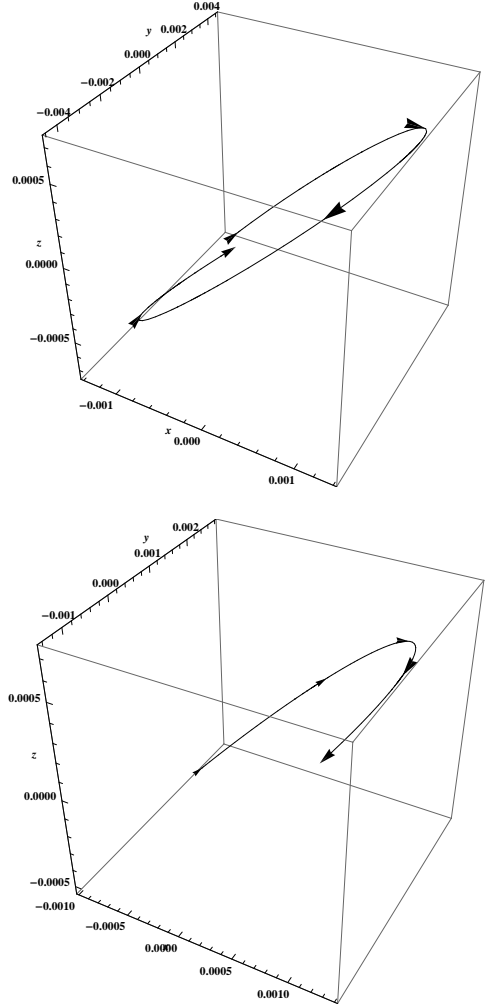}
 \caption{The effect of $\beta$ in one period} 
 \label{fig:lone}
 \end{figure} 
When the radiation pressure increases, the phase difference of the trajectories decrease. When there is no drag 
force (i.e. $\beta=0$), the trajectory of Lissajous orbit completes one period approximately at $t=3.0$, but when radiation 
pressure force increases (i.e. $\beta=0.2$), it does not complete its period at that time, as shown in Fig. \ref{fig:lone}.
Therefore, the period increases with an increase in the value of  $\beta$. Also, because of the increasing value of $\beta$, the 
trajectories shrink in its amplitude. 
\begin{figure}
\begin{center}
  \includegraphics[width=7cm]{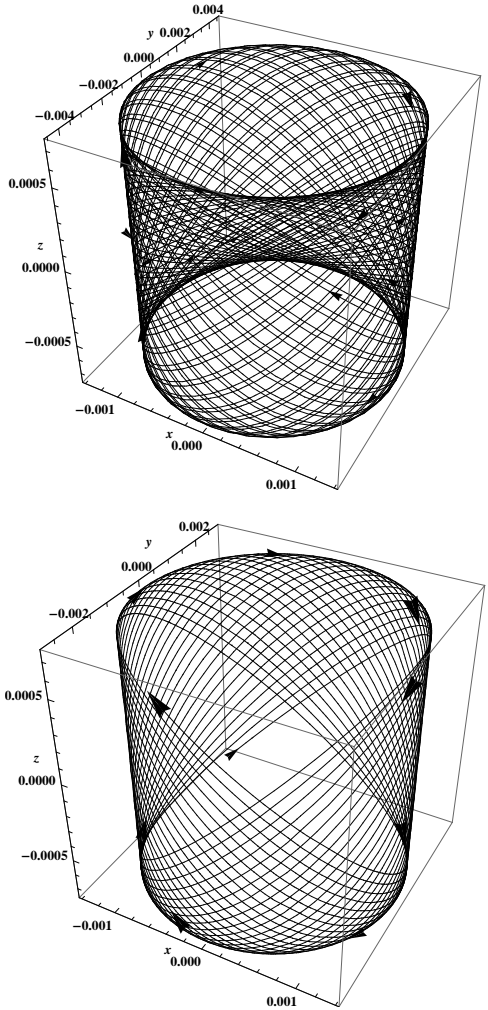}
 \caption{The effect of $\beta$ on Lissajous orbit} 
 \label{fig:lis}
 \end{center}
 \end{figure}
 \begin{figure}
\begin{center}
Figs.\ref{fig:lis} shows the Lissajous orbits with $\beta=0$ and $\beta=0.2$ shown in the upper and lower panels, respectively.
  \includegraphics[width=5cm]{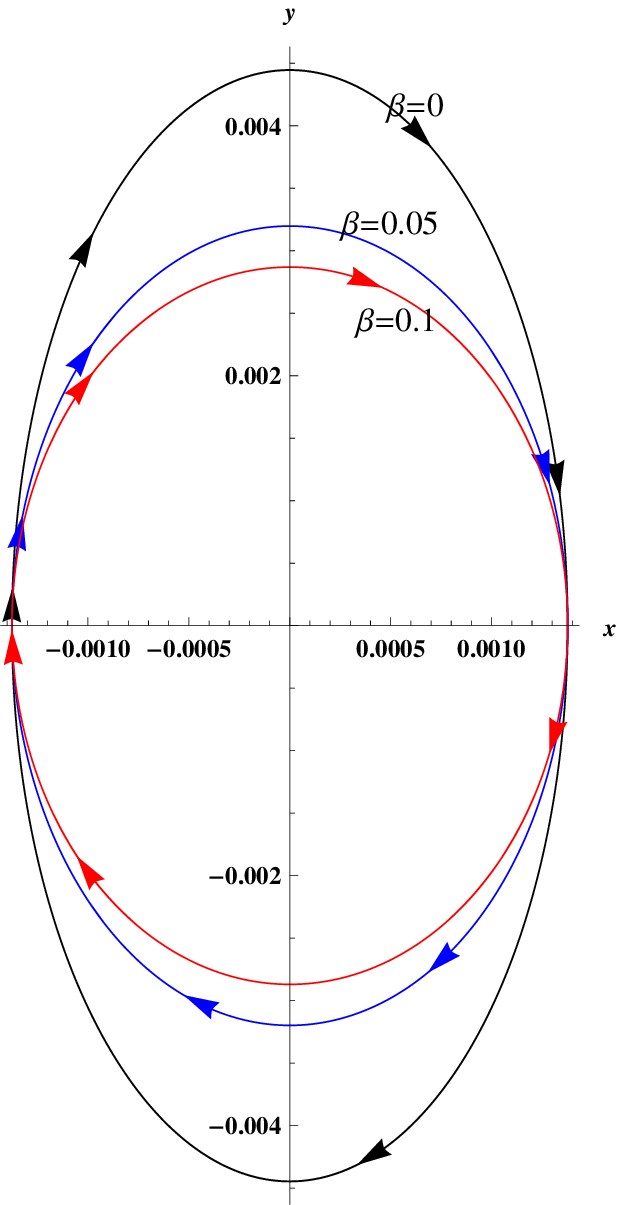}
 \caption{Projection on $xy-$plane} 
 \label{fig:lisa1}
 \end{center}
 \end{figure}
\begin{figure}
\begin{center}
  \includegraphics[height=7.5cm]{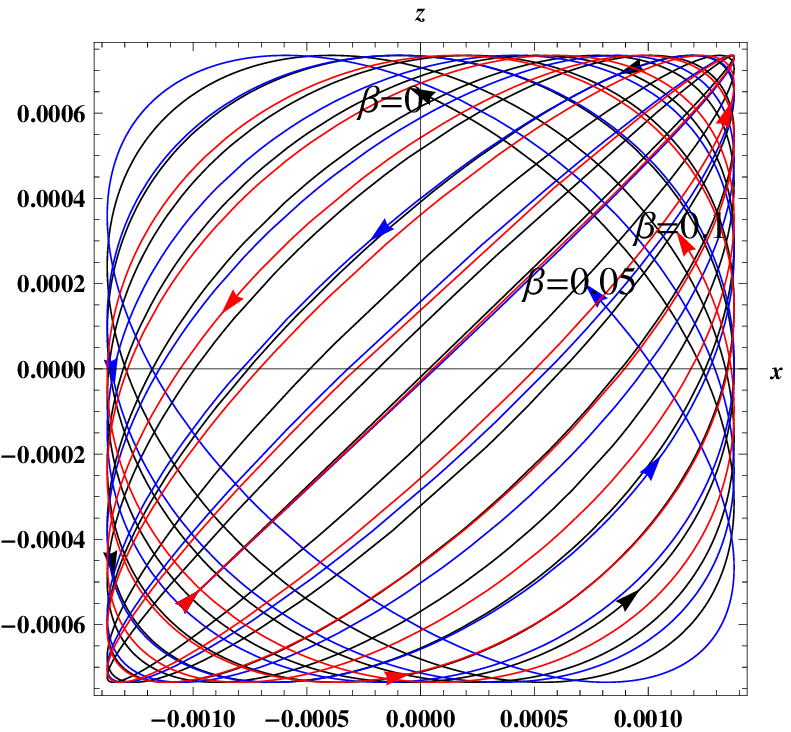}
 \caption{Projection on $xz-$plane} 
 \label{fig:lis1}
 \end{center}
 \end{figure}
 \begin{figure}
\begin{center}
  \includegraphics[height=7.5cm]{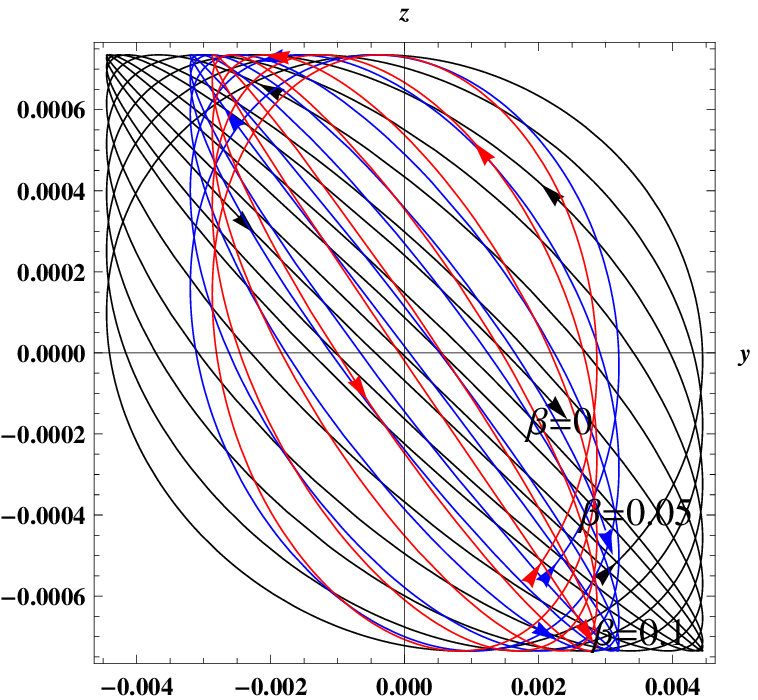}
 \caption{Projection on $yz-$plane} 
 \label{fig:lisa2}
 \end{center}
 \end{figure}  
 Figs. \ref{fig:lisa1}, \ref{fig:lis1} and \ref{fig:lisa2} show the projections in the  $xy$-, $xz$- and $yz$-plane, respectively.
  with the different values of $\beta$. Here, black, blue and red coloured orbits are shown with $\beta=0$, $0.05$ and $0.1$,  
  respectively. Clearly, the orbits shrink when $\beta$ increases.
 \subsection{Periodic orbits using the Lindstedt-Poincar$\acute{e}$ method}
 The equations of motions can be written using the Legendre polynomials $P_{n}$. A third-order approximation was 
 used in the circular restricted three-body without any drag force by \cite{Richardson1980CeMec..22..241R}. Here, we include 
 P-R drag and solar wind drag, and then find the third-order approximation as described by \cite{Thurman96thegeometry}:
 \begin{eqnarray}
  \ddot{x}-2\dot{y}&=&x\nu_{10}-y\nu_{11}+x^2\nu_{12}+y^2 \nu_{13}+z^2\nu_{14}-\nonumber\\&&
  xy\nu_{15}+x^3 \nu_{16}+y^3 \nu_{17}+xy^2 \nu_{18}+xz^2 \nu_{19}+\nonumber\\&& x^2 y\nu_{190}+yz^2 \nu_{191}+\nu^{*}+O(4),
   \end{eqnarray}
  \begin{eqnarray}
  \ddot{y}+2\dot{x}&=&x\nu_{20}+y\nu_{21}+x^2\nu_{22}+y^2\nu_{23}+z^2 \nu_{24}+\nonumber\\&&
  xy\nu_{25}+x^3 \nu_{26}+y^3 \nu_{27}+xy^2 \nu_{29}+xz^2 \nu_{290}+\nonumber\\&& x^2 y\nu_{291}+yz^2\nu_{292}+\nu_{**}+O(4),
   \end{eqnarray}
  \begin{eqnarray}
 \ddot{z}+\nu_{*}z&=&xz\nu_{30}+yz\nu_{31}+ x^2 z\nu_{32}+x y z \nu_{33} +\nonumber\\&&y^2 z\nu_{34}+z^3 \nu_{35}+O(4).\label{eq:z}
 \end{eqnarray}
All the coefficients are given in Appendix A.

\subsection{Correction term}
For the construction of higher-order applications, we linearize equation (\ref{eq:z}) and introduce a correction term
 \begin{eqnarray}
  &&\Delta=\lambda^2-\nu_{*},
 \end{eqnarray}
on the right-hand side. The new third-order z equation becomes,
\begin{eqnarray}
 \ddot{z}+\lambda^2 z&=&xz\nu_{30}+yz\nu_{31}+x^2 z\nu_{32}+xyz \nu_{33}\nonumber\\&& +y^2 z \nu_{34}+z^3 \nu_{35}+
 \Delta z+O(4).
\end{eqnarray}
\cite{Richardson1980CeMec..22..241R} developed a third-order periodic solution using the Lindstedt-Poincar$\acute{e}$ 
type of successive approximations. We follow their work with the P-R drag and solar wind drag, by removing secular terms. A
new independent variable $\tau$ and a frequency connection $\omega$ are introduced via, $\tau=\omega t$.
 Then, the equations of motion at the second degree including the P-R drag and solar wind drag are given as
\begin{eqnarray}
 \omega^2 x''-2\omega y'&=&x\nu_{10}-y\nu_{11}+x^2\nu_{12}+y^2 \nu_{13}+z^2\nu_{14}\nonumber\\&&- xy\nu_{15}+x^3 \nu_{16}+
y^3 \nu_{17}+xy^2 \nu_{18}+xz^2 \nu_{19}\nonumber\\&&+x^2 y\nu_{190}+yz^2 \nu_{191}+\nu^{*}+O(4)\label{eq:ram1},
\end{eqnarray}
\begin{eqnarray}
 \omega^2 y''+2\omega x'&=&x\nu_{20}+y\nu_{21}+x^2\nu_{22}+y^2\nu_{23}+z^2 \nu_{24}\nonumber\\&&+
  xy\nu_{25}+x^3 \nu_{26}+y^3 \nu_{27}+xy^2 \nu_{29}+\nonumber\\&&xz^2 \nu_{290}+ x^2 y\nu_{291}+yz^2\nu_{292}\nonumber\\&&+
  \nu_{**}+O(4),
  \end{eqnarray}
  \begin{eqnarray}
  \omega^2 z''+\lambda^2 z&=&xz\nu_{30}+yz\nu_{31}+xy\nu_{32}+x^2 \nu_{33}\nonumber\\&&+
  x^2 z\nu_{32}+x y z \nu_{33} +y^2 z\nu_{34}+z^3 \nu_{35}\nonumber\\&&+
 \Delta z+O(4),
\end{eqnarray}
where  a prime denotes $\frac{d}{d\tau}$. Because of perturbation analysis, we assume solutions in the following form:
\begin{eqnarray}
 &&x(\tau)=\epsilon x_{1}(\tau)+\epsilon^2 x_{2}+\epsilon^3 x_{3}\dots;
 \end{eqnarray}
  \begin{eqnarray}
 &&y(\tau)=\epsilon y_{1}(\tau)+\epsilon^2 y_{2}+\epsilon^3 y_{3}\dots;
 \end{eqnarray}
  \begin{eqnarray}
 &&z(\tau)=\epsilon z_{1}(\tau)+\epsilon^2 z_{2}+\epsilon^3 z_{3}\dots;
\end{eqnarray}
\begin{eqnarray}
 &&\omega=1+\epsilon \omega_{1}(\tau)+\epsilon^2 \omega_{2}+\epsilon^3 \omega_{3}\dots .\label{eq:ram2}.
\end{eqnarray}
We substitute these into the equations of motion and equate components of the same order of $\epsilon$. We find the first-, 
second-, and third-order equations and their solutions.

\subsection{First-order equations}
To find the first-order approximation, we use equations (\ref{eq:ram1})-(\ref{eq:ram2}) and comparing the coefficients of the 
first-order terms in $\epsilon$ on both sides. We obtain the following equations:
\begin{eqnarray}
  x''-2 y'&=&x\nu_{10}-y\nu_{11}+x^2\nu_{12};
 \end{eqnarray}
  \begin{eqnarray}
 y''+2 x'&=&x\nu_{20}+y\nu_{21};
 \end{eqnarray}
  \begin{eqnarray}
 z''+\lambda^2 z&=&0.
\end{eqnarray}
From above equations, we find the following bounded solutions:
\begin{eqnarray}
 x_{1}(\tau)&=&-A_{x} \cos(\lambda \tau +\phi);
 \end{eqnarray}
  \begin{eqnarray}
y_{1}(\tau)&=&\kappa A_{x}\{2\lambda \sin(\lambda\tau+\phi)-\nu_{10}\cos(\lambda \tau +\phi)\};
 \end{eqnarray}
  \begin{eqnarray}
 z_{1}(\tau)&=&A_{z}\sin(\lambda \tau +\psi).
\end{eqnarray}
In order to avoid secular solutions, we need to constraints on the constants $A_{x}$, $A_{z}$, $\phi$ and $\psi$, but 
for now they are arbitrary.
\subsection{Second-order equations}
The order of $O(\epsilon^2)$ equations depend on the first order solutions for $x_1, y_1,z_1$. Collecting only non-secular 
terms, we obtain
\begin{eqnarray}
x''_{2}-2y'_{2}-\nu_{10} x_{2}+\nu_{11}y_{2}&=&\alpha_{1}\cos2\tau_{1}-\alpha_{2}\cos2\tau_{2}\nonumber\\&&-
 \alpha_{3}\sin2\tau_{1}+\alpha_{4},
 \end{eqnarray}
  \begin{eqnarray}
  y''_{2}+2x'_{2}-\nu_{21}y_{2}-\nu_{20}x_{2}&=&-\delta_{1}\sin2\tau_{1}+\delta_{2}\cos2\tau_{1}\nonumber\\&&-
  \delta_{3}\cos2\tau_{2}+\delta_{4},
  \end{eqnarray}
  \begin{eqnarray}
  z''_{2}+\lambda^2 z_{2}&=&-h_{1}[\sin(\tau_{1}+\tau_{2})+\sin(\tau_{2}-\tau_{1})]\nonumber\\&&+
h_{2}[\cos(\tau_{2}-\tau_{1})-\cos(\tau_{1}+\tau_{2})],
\end{eqnarray}
where all the coefficients are given in Appendix A.

We remove all the secular terms by setting $\omega_{1}=0$. Thus, we find the following solutions, 
\begin{eqnarray}
x_{2}&=&\rho_{10}+\rho_{11}\cos2\tau_{1}+\rho_{12}\cos2\tau_{2}+\nonumber\\&&\rho_{13}\sin2\tau_{2}+\rho_{14}\sin2\tau_{1},
\end{eqnarray}
  \begin{eqnarray}
y_{2}&=&\rho_{20}+\rho_{21}\sin2\tau_{1}+\rho_{22}\sin2\tau_{2}-\nonumber\\&&\rho_{23}\cos2\tau_{2}-\rho_{24}\cos2\tau_{1},
\end{eqnarray}
  \begin{eqnarray}
 z_{2}&=&\rho_{30}\sin(\tau_{1}+\tau_{2})+\rho_{31}\sin(\tau_{2}-\tau_{1})\nonumber\\&&+\rho_{32}\cos(\tau_{2}-\tau_{1})+
 \rho_{33}\cos(\tau_{1}+\tau_{2}),
\end{eqnarray} 
where
\begin{eqnarray}
 &&\tau_{1}=\lambda\tau+\phi, \ \ \ \mbox{and} \ \ \ \tau_{2}=\lambda\tau+\psi.
\end{eqnarray}
All the coefficients are given in Appendix A.
\subsection{Third-order equations}
The $O(\epsilon^3)$ equations are obtained by setting $\omega_1 =0$ and 
substituting in the solutions for $x_{1}, y_{1}, z_{1}, x_{2}, y_{2}$ and $z_{2}$. Thus we obtain 
\begin{eqnarray}
 &&x_{3}''-2 y_{3}'-\nu_{10}x_{3}+\nu_{11}y_{3}=[\alpha_{11}+\nonumber\\&&2\omega_{2}\lambda^2 A_{x}(2\kappa-1)]
 \cos\tau_{1}+[\alpha_{12}+\nonumber\\&&2\kappa \lambda A_{x}\nu_{11}^{2}\omega_{2}]\sin \tau_{1}+\alpha_{13}\cos 3\tau_{1}+
\nonumber\\&& \alpha_{14}\cos(\tau_{1}+2\tau_{2})+\alpha_{15}\cos(\tau_{1}-2\tau_{2})+\nonumber\\&&\alpha_{16}\sin(2\tau_{2}+
\tau_{1})+\alpha_{17}\sin(2\tau_{2}-\tau_{1})+\nonumber\\&&\alpha_{18}\sin3\tau_1,\label{eq:ree1}
\end{eqnarray}
\begin{eqnarray}
&&y_{3}''+2x_{3}'-\nu_{20}x_{3}-\nu_{21}y_{3}=[\alpha_{21}-\nonumber\\&&2\kappa\lambda^2 A_{x}\nu_{11}\omega_{2}]\cos\tau_{1}
 +[\alpha_{22}+\nonumber\\&&2\lambda A_{x}\omega_{2}(2\kappa\lambda^2 -1)]\sin\tau_1+\nonumber\\&&
 \alpha_{23}\cos3\tau_{1}+\alpha_{24} \cos(\tau_{1}+2\tau_{2})+\nonumber\\&&\alpha_{25}\cos(2\tau_{2}-\tau_1)+
 \alpha_{26}\sin(2\tau_{2}+\tau_{1})+\nonumber\\&&\alpha_{27}\sin(2\tau_{2}-\tau_{1})+\alpha_{28}\sin3\tau_1,\label{eq:ree2}
\end{eqnarray}
\begin{eqnarray}
 &&z_{3}''+\lambda^2 z_{3}=\left[\alpha_{31}+A_{z}\left(2\omega_{2}\lambda^2 +\frac{\Delta}{\epsilon^2}\right)\right]\sin\tau_{2}+
 \nonumber\\&&\alpha_{32} \sin(2\tau_{1}+\tau_{2})+\alpha_{33}\sin(\tau_{2}-2\tau_{1})+\nonumber\\&&
 \alpha_{34}\sin3\tau_2 +\alpha_{35}\cos\tau_2 + \alpha_{36}\cos3\tau_2 +\nonumber\\&&\alpha_{37}\cos(\tau_{2}-2\tau_1)+
 \alpha_{38}\cos(2\tau_1 +\tau_2),
\end{eqnarray}
where the expressions of the coefficients are given in Appendix A. There are secular terms. We start by examining the secular 
terms in the $z_{3}$ equation, by simply setting a value for the frequency correction $\omega_{2}$. To remove the secular 
terms $\alpha_{33}\sin(\tau_2 -2\tau_1)$ and $\alpha_{37}\cos(\tau_2 -2\tau_1)$, we need the coefficients of these terms 
to be zero. Thus
\begin{eqnarray}
 &&\phi=\psi+n\frac{\pi}{2}, \ \ \mbox{where} \  \ n=0,1,2,3. \nonumber
\end{eqnarray}
The solution will be bounded if 
\begin{eqnarray}
 &&\alpha_{31}+A_{z}\left(2\omega_{2}\lambda^2 +\frac{\Delta}{\epsilon^2}\right)-\zeta\alpha_{33}=0\label{eq:al},\\ \ \mbox{and} 
 \ &&\zeta\alpha_{37}=0,
\end{eqnarray}
where $\zeta=(-1)^n$. This phase constraint affects  the $x_{3}-y_{3}$ equations; now each contains a secular term. The 
requirement of another constraint is from the simultaneous equations (\ref{eq:ree1}) and (\ref{eq:ree2}):
\begin{eqnarray}
&&-(\nu_{11}+\lambda^2)[\alpha_{11}+2\omega_{2}\lambda^2 A_{x}(2\kappa-1)]+\nonumber\\&&+2\lambda[\alpha_{22}+
2\lambda A_{x}\omega_{2}(2\kappa \lambda^2 -1)]+\nonumber\\&&\nu_{11}(\alpha_{21}-2\kappa\lambda^2 A_{x}\nu_{11}\omega_{2})+
\zeta[-\alpha_{15}(\lambda^2 +\nonumber\\&&\nu_{21})+\lambda(\alpha_{27}+\nu_{11}\alpha_{25})]=0,
\end{eqnarray}
and
\begin{eqnarray}
 &&-(\nu_{22}+\lambda^2)(\alpha_{12}+2\kappa\lambda A_{x}\nu_{11}^2 \omega_{2})-\nonumber\\&&2\lambda(\alpha_{21}-2\kappa\lambda^2 A_{x}
 \nu_{11}\omega_{2})+\nu_{11}[\alpha_{22}+\nonumber\\&&2\lambda A_{x}\omega_{2}(2\kappa\lambda^2 -1)]+\zeta[-\alpha_{17}(\lambda^2 +
 \nu_{21})+\nonumber\\&&\lambda(\alpha_{25}+\nu_{11}\alpha_{27})]=0.
\end{eqnarray}
From these equations we find 
\begin{eqnarray}
 &&\omega_{2}=[\lambda^2 \alpha_{11}+\zeta \lambda^2 \alpha_{15}-2\lambda\alpha_{22}-\zeta\lambda\alpha_{27}+\nonumber\\&&
 \alpha_{11}\nu_{11}-\alpha_{21}\nu_{11}-\zeta(\lambda\alpha_{25}\nu_{11}-\alpha_{15}\nu_{21})] \nonumber\\&&
 \{2\lambda^2 A_{x}\{\lambda^2 (2\kappa+1)+\nu_{11}-2\kappa\nu_{11}-\kappa\nu_{11}^2 -2]\}^{-1}\label{eq:om}.
 \end{eqnarray}
The amplitude relation is obtained by putting the value of equation (\ref{eq:om}) into equation (\ref{eq:al}). Thus, we can 
satisfy this constraint by letting one amplitude be determined by the other.
\begin{table*}
  \centering
  \begin{minipage}{140mm}
   \caption{At $\beta=0.13$}\label{tbl:3}
   \begin{tabular}{@{}cccccc@{}}
   \hline
$n$&$A_{z}$& $\omega_{2}$\\
\hline
0 & $\pm0.4202186$ &$-24.7175$\\
1 & $\pm0.1120075$ &$-4.66144$ \\
2 & $\pm0.4202186$ &$-24.7175$\\
3 & $\pm0.1120075$ &$-4.66144$\\
\hline
 \end{tabular}
\end{minipage}
 \end{table*}  
\begin{table*}
  \centering
  \begin{minipage}{140mm}
   \caption{At $\beta=0.15$}\label{tbl:a3}
   \begin{tabular}{@{}cccccc@{}}
   \hline
$n$&$A_{z}$& $\omega_{2}$\\
\hline
0 & $\pm0.219213958$ &$-5.14832$\\
1 & $\pm0.105566928$ &$-3.32111$\\
2 & $\pm0.219213958$ &$-5.14832$\\
3 & $\pm0.105566928$ &$-3.32111$\\
\hline
 \end{tabular}
\end{minipage}
 \end{table*}
 \begin{table*}
  \centering
  \begin{minipage}{140mm}
   \caption{At $\beta=0.18$}\label{tbl:a4}
   \begin{tabular}{@{}cccccc@{}}
   \hline
$n$&$A_{z}$& $\omega_{2}$\\
\hline
0 & $\pm0.09654386$ &$-0.718982$\\
1 & $\pm0.05566177$ &$-0.691666$\\
2 & $\pm0.09654386$ &$-0.718982$\\
3 & $\pm0.05566177$ &$-0.691666$\\
\hline
 \end{tabular}
\end{minipage}
 \end{table*}
 
Different amplitude $A_{z}$ at $A_{x}=206000/(1.496\times 10^8)$ with $sw=0.35$ at different $\beta$ and 
 $n$ are shown in Table \ref{tbl:3}, \ref{tbl:a3} and \ref{tbl:a4}. From these tables, we see that the amplitudes contract, 
 whereas $\omega_{2}$ decreases negatively when  $\beta$  increases. Using these constraints, the third-order equations are 
 reduced to  
\begin{eqnarray}
 &&x_{3}''-2 y_{3}'-\nu_{10}x_{3}+\nu_{11}y_{3}=s_{11} \cos\tau_{1} +\nonumber\\&&s_{12} \sin \tau_{1} +
 (\alpha_{13}+\zeta\alpha_{14})\cos 3 \tau_{1} +\nonumber\\&&(\alpha_{18}+\zeta\alpha_{16}) \sin 3\tau_{1},
\end{eqnarray}
\begin{eqnarray}
 &&y_{3}''+2x_{3}'-\nu_{20}x_{3}-\nu_{21}y_{3}=s_{21} \cos \tau_{1}+\nonumber\\&&s_{22} \sin \tau_{1}+(\alpha_{23}+
 \zeta\alpha_{24})\cos 3\tau_{1}+\nonumber\\&&(\alpha_{24}+\zeta\alpha_{26})\sin 3 \tau_{1},
\end{eqnarray}
\begin{eqnarray}
 z_{3}''+\lambda^{2}z_{3}&=&\zeta(\alpha_{32} \sin3\tau_{1}+\alpha_{38} \cos3\tau_{1})+\nonumber\\&&
   \alpha_{34}
\begin{cases}
    (-1)^{\frac{n}{2}} \sin 3\tau_{1},&  n=0, 2\\
    (-1)^{\frac{(n-1)}{2}} \cos 3\tau_{1}, &  n=1,3 \nonumber 
\end{cases}\\&& +\alpha_{36}
\begin{cases}
 (-1)^{\frac{n}{2}} \cos3\tau_{1}, &  n=0,2\\
 (-1)^{\frac{(n+1)}{2}} \sin 3 \tau_{1}, & n=1,3.
\end{cases}
\end{eqnarray}
The solutions are 
\begin{eqnarray}
 x_{3}&=&\sigma_{10}\cos\tau_1 +\sigma_{11}\sin\tau_1+\sigma_{12}\cos3\tau_1+\nonumber\\&&\sigma_{13}\sin3\tau_1,\\
 y_{3}&=&\sigma_{20}\cos\tau_1+\sigma_{21}\sin\tau_1+\sigma_{22}\cos3\tau_1+\nonumber\\&&\sigma_{23}\sin3\tau_1,\\
 z_{3}&=&\frac{1}{8\lambda^2}\left[\zeta(\alpha_{32}\sin3\tau_1+\alpha_{38}\cos3\tau_1)+\right.\nonumber\\&&\left.
 \begin{cases}
 (-1)^{\frac{n}{2}} (\alpha_{34}\sin3\tau_{1}+\alpha_{36}\cos3\tau_1),  n=0,2\\
 (-1)^{\frac{(n-1)}{2}} (\alpha_{34}\cos 3 \tau_{1}-\alpha_{36}\sin3\tau_1), n=1,3
\end{cases}\right],\nonumber\\&&
\end{eqnarray}
where the expressions of the coefficients are given in Appendix A.

\subsection{The final approximation}
Using all the first-, second- and third-order approximate solutions with the mapping $A_{x}\mapsto A_{x}/\epsilon$ 
and $A_{z}\mapsto A_{z}/\epsilon$, which remove $\epsilon$ from all the 
equations, we obtain 
\begin{eqnarray}
x(\tau)&=&\rho_{10}+(-A_{x}+\sigma_{10})\cos\tau_{1}+\sigma_{11}\sin\tau_1\nonumber\\&&+(\rho_{11}+
 \zeta\rho_{12})\cos2\tau_{1}+ (\rho_{14}+\zeta\rho_{13})\sin2\tau_{1}\nonumber\\&&+\sigma_{12}\cos3\tau_1 +
 \sigma_{13}\sin3\tau_1,
 \end{eqnarray} 
 \begin{eqnarray} 
 y(\tau)&=&\rho_{20}+(2\lambda\kappa A_{x}+\sigma_{21})\sin\tau_{1}\nonumber\\&&+(-\kappa A_{x}\nu_{11}+
\sigma_{20})\cos\tau_{1}+(\rho_{21}+ \zeta\rho_{22})\sin2\tau_{1}\nonumber\\&&-(\rho_{24}+\zeta\rho_{23})\cos2\tau_{1}+
\sigma_{22}\cos3\tau_1 \nonumber\\&&+\sigma_{23}\sin3\tau_1,
\end{eqnarray}
and
\begin{eqnarray}
 z(\tau)&=&A_{z}
 \begin{cases}
 (-1)^{\frac{n+4}{2}}(A_{z}\sin \tau_1+\rho_{30}\sin2\tau_1), n=0,2 \\
 (-1)^{\frac{n+1}{2}}(A_{z}\cos \tau_1+\rho_{30}\cos2\tau_1), n=1,3
 \end{cases}\nonumber\\&&+\rho_{31}
 \begin{cases}
  0, n=0,2\\
  (-1)^{\frac{n+1}{2}}, n=1,3
 \end{cases}+\rho_{32}\nonumber
 \begin{cases}
  (-1)^{\frac{n}{2}}, n=0,2\\
  0, n=1,3
 \end{cases}\nonumber\\&&
  -\frac{1}{8\lambda^2}\left[\zeta(\alpha_{32}\sin3\tau_1+\alpha_{38}\cos3\tau_1)+\right.\nonumber\\&& \left.
 \begin{cases}
  (-1)^{n/2}(\alpha_{34}\sin3\tau_1+\alpha_{36}\cos3\tau_1),n=0,2\\
  (-1)^{n/2}(\alpha_{34}\cos3\tau_1-\alpha_{36}\sin3\tau_1), n=1,3
  \end{cases}\right].\nonumber\\&&
\end{eqnarray}
The expressions of the coefficients are given in Appendix A.

For the Sun-(Earth-Moon) $L_{1}$, the difference of both frequencies of the $z-$plane and $xy$-plane is quite small. The value of 
$A_{x}$ and $A_{z}$ are found using the relation of equations (\ref{eq:om}) and (\ref{eq:al}). 
\begin{figure}
\begin{center}
  \includegraphics{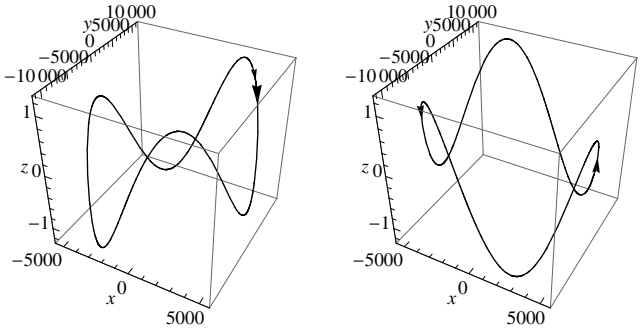}
 \caption{Halo orbit at $n=0$, and $n=2$  at $\beta=0.13$} 
 \label{fig:f0}
 \end{center}
 \end{figure}
\begin{figure}
\begin{center}
 \includegraphics{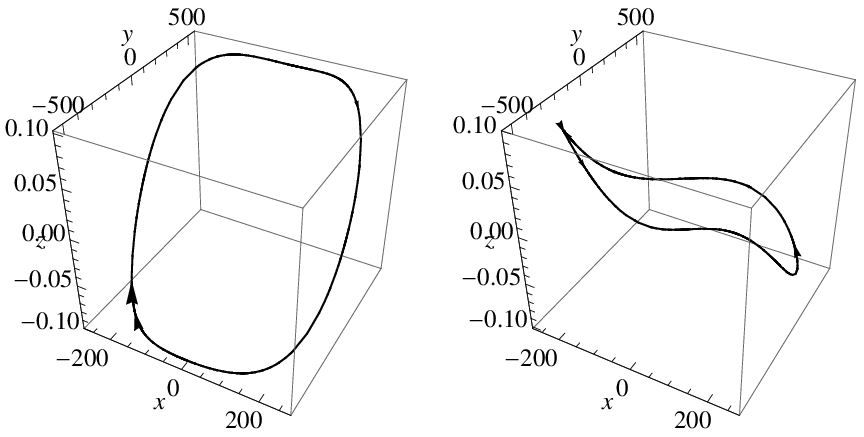}
 \caption{Halo orbit at $n=1$, and $n=3$ at $\beta=0.13$} 
 \label{fig:f1}
 \end{center}
 \end{figure} 
\begin{figure}
 \begin{center}
 \includegraphics{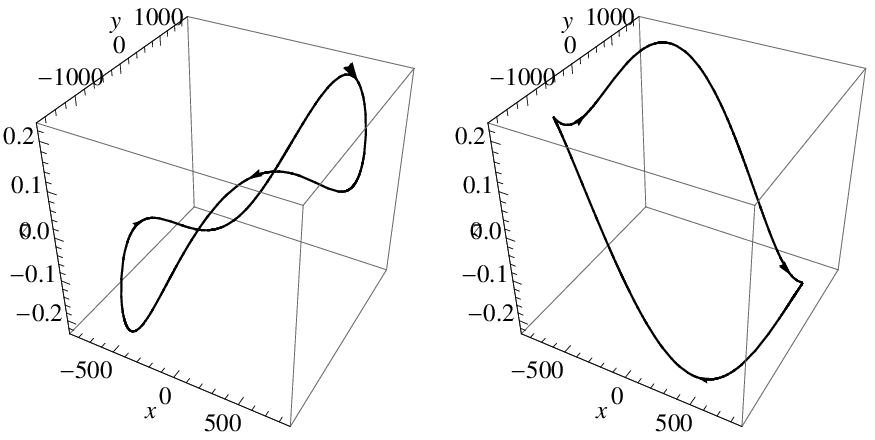}
 \caption{Halo orbit at $n=0$, and $n=2$ at $\beta=0.15$} 
 \label{fig:f2}
 \end{center}
 \end{figure}
 \begin{figure}
\begin{center}
 \includegraphics{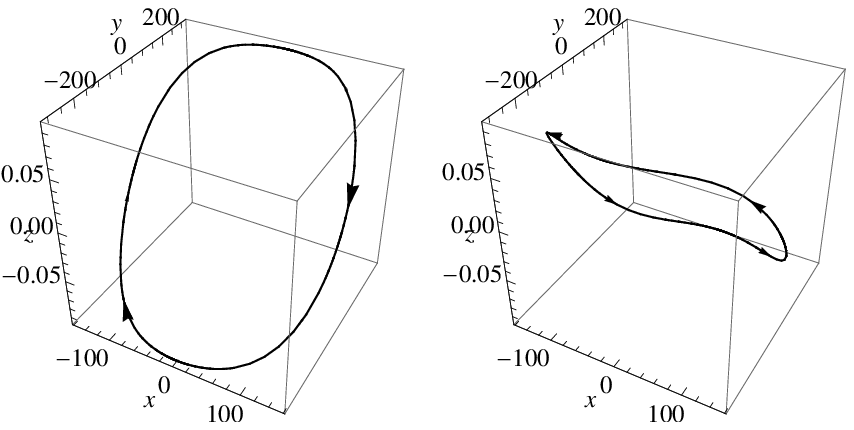}
 \caption{Halo orbit at $n=1$, and $n=3$ at $\beta=0.15$} 
 \label{fig:f3}
  \end{center}
 \end{figure}

 \begin{figure}
\begin{center}
 \includegraphics{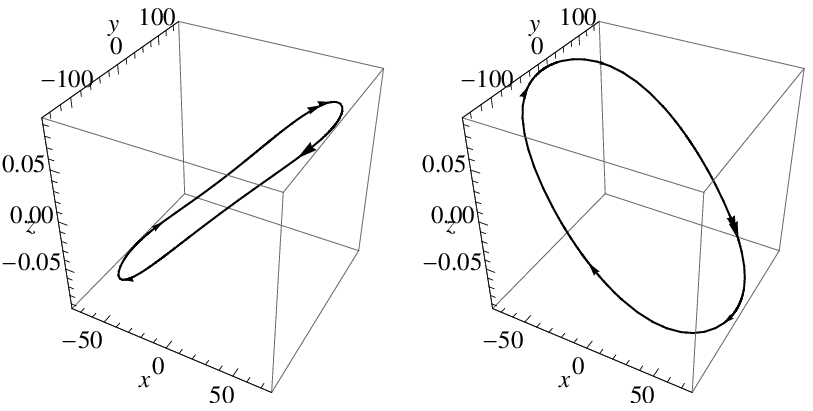}
 \caption{Halo orbit at $n=0$, and $n=2$ at $\beta=0.18$} 
 \label{fig:f4}
  \end{center}
 \end{figure}
 
 \begin{figure}
\begin{center}
 \includegraphics{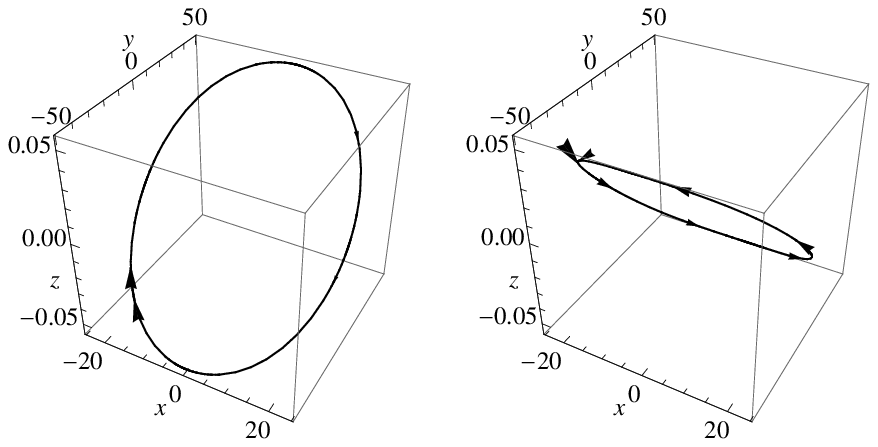}
 \caption{Halo orbit at $n=1$, and $n=3$ at $\beta=0.18$} 
 \label{fig:f5}
  \end{center}
 \end{figure}
 In the final approximation, the halo orbits at $n=0$ and $2$ with  $sw=0.35$ are shown in 
Figs. \ref{fig:f0}, \ref{fig:f2} and \ref{fig:f4}. For $n=1$ and $3$, the halo orbits are depicted in Figs \ref{fig:f1}, 
\ref{fig:f3} and  \ref{fig:f5}. The amplitude and frequency $\omega_{2}$ are both equal for each case, $n=0, \ 2$ and 
$n=1, \ 3$. When $\beta$ increases, then the  trajectory shrinks. 

In the previous subsections, we have computed the first, second, third and final approximations for the halo orbits and we have 
seen the effect of radiation pressure, with P-R drag and solar wind drag. In the absence of drag forces, the results agree with
those of the classical case \citep{Thurman96thegeometry}
 \section{Conclusions} 
In this paper, we have studied the circular restricted three-body problem of the Sun-Earth-Moon system by assuming the effect 
of radiation pressure, P-R drag and solar  wind drag. We have found that the collinear Lagrangian points deviate from their axis 
joining the primaries, 
whereas the triangular points remain unchanged in their configuration. However, all points lie in a plane. If we increase 
the value of $\beta$, with a fixed value of $sw=0.35$, the Lagrangian points $L_{1}$, $L_{2}$ and $L_{3}$ 
tend towards the radiating body (the Sun), whereas $L_{4,5}$ have symmetrical changes with the increasing value of  $\beta$. We 
have examined the linear stability of the equilibrium points with the help of characteristic roots. 
It is observed that the Lagrangian points are unstable because of the drag forces. Further, we have computed 
the orbit around the $L_{1}$ point and have seen that when radiation pressure increases, the phase difference of the 
trajectory decreases. If there is no drag force (i.e. $\beta=0$), then the trajectory of the Lissajous orbit completes one
period approximately at $t=3.0$, whereas if $\beta$ increases with $sw=0.35$, it does not complete its period at the same time. 
Also, because of the increasing value of $\beta$, the trajectories shrink in its amplitude. In this study, we have used the 
Lindstedt-Poincar$\acute{e}$ method to compute the halo orbits in the third-order approximation with the radiation pressure,
P-R drag and solar wind drag. In this analysis, we have fixed the value of $sw=0.35$, which is the ratio of 
of solar wind drag to P-R drag, and we have varied the value of $\beta$, (i.e. the ratio of radiation pressure 
force to solar gravitation force). This model can be used to compute the higher-order approximation (four, fifth, etc.)
expressions for halo orbits. Moreover, stable unstable manifolds of the halo orbits, and trajectory transfer would be  
interesting topics of future research with the similar dissipative forces.

\section*{Acknowledgments}
We are grateful to Inter-University Centre for Astronomy and Astrophysics (IUCAA), Pune for supporting 
library visits and for the use of computing facilities. BSK is also grateful to the Indian Space Research Organization (ISRO),
Department of Space, Government of India, for providing financial support through the RESPOND Programme 
(Project No.-ISRO/RES/2/383/2012-13).

 \bibliographystyle{mn2e} 

 \section{Appendix A: Coefficients}
 \begin{eqnarray}
 a_{00}&=&(a^*-d^*-1)(a^*- b^*-1)-c^{*2},\\\nonumber&&\\
 a_{20}&=&2(1+a^*)-b^*-d^*,\\\nonumber&&\\
  a_{3}&=&-(1+sw)\{K_{x,\dot{x}}+K_{y,\dot{y}}\},\\\nonumber&&\\
 a_2&=&-(1+sw)\{K_{x,x} +K_{y,y}-2(K_{x,\dot{y}}-K_{y,\dot{x}})\},\\\nonumber&&\\
 a_1&=&(1+sw)\{(1-a^*+b^*)K_{x,\dot{x}} +(1-a^*\nonumber\\&&+d^*)K_{y,\dot{y}}+ 2(K_{x,y}-K_{y,x})
 -c^*(K_{x,\dot{y}}+\nonumber\\&&K_{y,\dot{x}})\},\\\nonumber&&\\
 a_0&=&(1+sw)\{(1-a^*+b^*)K_{x,x}+(1-a^*+\nonumber\\&&d^*)K_{y,y}-c^*(K_{x,y}+K_{y,y})\},\\\nonumber&&\\
 \zeta_{1}&=&\nu_{20}\nu_{11}+\nu_{10}\nu_{21},\\\nonumber&&\\
 \nu_{10}&=&1-(c_{2}+d_{2})+3\left[\frac{c_{2}(\gamma-1)^2}{D_{1}^2}+\frac{d_{2}\gamma^2}{D_{2}^2}\right]+\nonumber\\&&
 \frac{(1+sw)\beta (1-\mu)}{c}\left[\frac{\mu l+2l(\gamma-1)}{D_{1}^4}-\nonumber\right.\\&&\left.
 \frac{4\mu(\gamma-1)^2 l}{D_{1}^6} \right],
 \end{eqnarray}
 \begin{eqnarray}
 \nu_{11}&=&3l\left[\frac{c_{2}(\gamma-1)}{D_{1}^2}+\frac{d_{2}\gamma}{D_{2}^2}\right]+\nonumber\\&&\frac{(1+sw)\beta 
 (1-\mu)}{c} \left[\frac{\mu(\gamma-1)}{D_{1}^4}-\frac{1}{D_{1}^2}\right],\\\nonumber&&\\
 \nu_{12}&=&\frac{15}{2}\left[\frac{c_{3}(\gamma-1)^3}{D_{1}^3}+\frac{d_{3}\gamma^3}{D_{2}^3}\right]-\nonumber\\&&
 \frac{9}{2} \left[\frac{c_{3}(\gamma-1}{D_{1}}+\frac{d_{3}\gamma}{D_{2}}\right]+\frac{(1+sw)\beta (1-\mu)}{c}\times\nonumber
 \\&& \left[\frac{6\mu(\gamma-1)l+4l(\gamma-1)}{D_{1}^6}-\frac{l}{D_{1}^4}-
 \frac{12\mu l(\gamma-1)^3}{D_{1}^8}\right]\nonumber,\\&&\\\nonumber&&\\
 \nu_{13}&=&-\frac{3}{2}\left[\frac{c_{3}(\gamma-1)}{D_{1}}+\frac{d_{3}\gamma}{D_{2}}\right]+\nonumber\\
 &&\frac{(1+sw)\beta(1-\mu)}{c} \left[ \frac{4\mu(\gamma-1)l}{D_{1}^{6}}-\frac{2l}{D_{1}^4}\right],\\\nonumber&&\\
 \nu_{14}&=&\frac{-3}{2}\left[\frac{c_{3}(\gamma-1)}{D_{1}}+\frac{d_{3}\gamma}{D_2}\right]+\nonumber\\&&
 \frac{(1+sw)\beta(1-\mu)}{c} \left[\frac{2\mu l(\gamma -1)}{D_{1}^6}-\frac{l}{D_{1}^4}\right],\\\nonumber&&\\
 \nu_{17}&=&\frac{15l}{2}\left[\frac{c_{4}(\gamma-1)}{D_{1}^2}+\frac{d_{4}\gamma}{D_{2}^2}\right]\nonumber\\&&
 \frac{(1+sw)\beta(1-\mu)}{c}\left[\frac{2\mu(\gamma-1)}{D_{1}^6}-\frac{1}{D_{1}^4}\right], \\\nonumber&&\\
 \nu_{15}&=&15 l\left[\frac{c_{3}(\gamma-1)^2}{D_{1}^3}+\frac{d_{3}\gamma^2 }{D_{2}^3}\right]-\nonumber\\&&
 3l\left[\frac{c_3}{D_1}+ \frac{d_{3}}{D_{2}}\right]+\frac{105 l}{2}\left[\frac{c_{4}(\gamma-1)^3}{D_{1}^4}+
 \frac{d_{4}\gamma^3}{D_{2}^4}\right]\nonumber\\&&-\frac{(1+sw)\beta (1-\mu)}{c}
 \left[\frac{\mu+2(\gamma-1)}{D_{1}^4}-\right.\nonumber\\&&\left. \frac{4\mu(\gamma-1)^2}{D_{1}^6}\right),\\\nonumber&&\\
  \nu_{16}&=&-15\left[\frac{c_{4}(\gamma-1)^2}{D_{1}^2}+\frac{d_{4}\gamma^2}{D_{2}^2}\right]+\nonumber\\&&
  \frac{35}{2}\left[\frac{c_{4}(\gamma-1)^{4}}{D_{1}^{4}}+\frac{d_{4}(\gamma-1)^2}{D_{2}^4}\right]+
  \frac{3}{2}(c_{4}+d_{4})+\nonumber\\&&\frac{(1+sw) \beta(1-\mu)}{c}\left[
  \frac{24l\mu(\gamma-1)^2 +8l(\gamma-1)^3}{D_{1}^8}-\right.\nonumber\\&&\left.
  \frac{28((\gamma-1)^4 l \mu)}{D_{1}^{10}}-\frac{2\mu l+4(\gamma-1)l}{D_{1}^6}\right],\\\nonumber&&\\
 \nu_{18}&=&-\frac{15}{2}\left[\frac{c_{4}(\gamma-1)^2}{D_{1}^2}+\frac{d_{4}\gamma^2}{D_{2}^2}\right]+
 \frac{3}{2}\left(c_{4}+d_{4}\right)\nonumber\\&&+
 \frac{(1+sw)\beta(1-\mu)}{c}\left[\frac{-6\mu l-12 l(\gamma-1)}{D_{1}^6}+\right.\nonumber\\&&\left.
 \frac{(24\mu l-12\mu)(\gamma-1)^2}{D_{1}^8}\right],\\\nonumber&&\\
 \nu_{19}&=&\frac{-15}{2}\left[\frac{c_{4}(\gamma-1)^2}{D_{1}^2}+\frac{d_{4}\gamma^2}{D_{2}^2}\right]+
 \frac{3}{2}(c_{4}+d_{4})\nonumber\\&&
 +\frac{(1+sw)\beta(1-\mu)}{c}\left[-\frac{2\mu l+4(\gamma-1)l}{D_{1}^6}-
 \frac{12\mu(\gamma-1)^2 l}{D_{1}^8}\right],\nonumber\\
 \end{eqnarray}
 \begin{eqnarray}
 \nu^{*}&=&1-\mu-\gamma +\frac{c_{1}(\gamma-1)}{D_{1}}+\frac{d_{1}\gamma}{D_{2}}+\nonumber\\&&\frac{(1+sw)\beta(1-\mu)}{c}
 \left[\frac{l}{D_{1}^2}-\frac{\mu l (\gamma-1)}{D_{1}^4}\right],\\ \nonumber&&\\
  \nu_{20}&=&-3l\left[\frac{c_{2}(\gamma-1)}{D_{1}^2}+\frac{d_{2}\gamma}{D_{2}^2}\right]-\nonumber\\&&\frac{(1+sw)
  \beta (1-\mu)}{c} \left[\frac{2(\gamma-1)(1-\mu-\gamma)}{D_{1}^4}+\frac{1}{D_{1}^2}\right],\nonumber\\&&\\\nonumber&&\\
 \nu_{21}&=&1-c_{2}-d_{2}-\frac{2(1+sw)\beta (1-\mu)(\gamma-1)l}{cD_{1}^4}, \\\nonumber&&\\
 \nu_{22}&=&-\frac{15}{2}l\left[\frac{c_{3}(\gamma-1)^2}{D_{1}^3}+\frac{d_{3}\gamma^2}{D_{2}^3}\right]+\nonumber\\&&
\frac{3l}{2}\left[ \frac{c_{3}}{D_1}+\frac{d_{3}}{D_{2}}\right]+\frac{(1+sw)\beta (1-\mu)}{c}\times\nonumber\\&&\left[
\frac{3(1-\gamma)-\mu}{D_{1}^4}-\frac{4(\gamma-1)^2 (1-\mu-\gamma)}{D_{1}^6}\right],\\\nonumber&&\\
\nu_{23}&=&\frac{9l}{2}\left[\frac{c_{3}}{D_{1}}+\frac{d_{3}}{D_{2}}\right]-\frac{(1+sw)\beta(1-\mu)}{c}
 \left[\frac{(\gamma-1)}{D_{1}^4} \right],\nonumber\\&&\\\nonumber&&\\
\nu_{24}&=&\frac{3l}{2}\left[\frac{c_{3}}{D_{1}}+ \frac{d_{3}}{D_{2}}\right]+
 \frac{(1+sw)\beta(1-\mu)(1-\mu-\gamma)}{cD_{1}^4},\nonumber\\&&\\\nonumber&&\\
 \nu_{25}&=&-3\left[\frac{c_{3}(\gamma-1)}{D_1}+\frac{d_{3}\gamma}{D_2}\right]+\frac{(1+sw)\beta (1-\mu)}{c}\times\nonumber\\&&
 \left[\frac{8 l(\gamma-1)^{2}l\{-3(\mu+\gamma)+2\}}{D_{1}^6} +\frac{2l}{D_{1}^4}\right],\\\nonumber&&\\
 \nu_{30}&=&-3\left[\frac{c_{3}(\gamma-1)}{D_{1}}+\frac{d_{3}\gamma}{D_2}\right]+\nonumber\\&&
 \frac{4(1+sw)\beta \mu(1-\mu)(\gamma-1)l}{cD_{1}^6},\\\nonumber&&\\
 \nu_{31}&=&3l\left[\frac{c_{3}}{D_{1}}+\frac{d_{3}}{D_{2}}\right]+\frac{(1+sw)\beta \mu(1-\mu)}{cD_{1}^4},\\\nonumber&&\\
 \nu_{190}&=&\frac{45l}{2}\left[\frac{c_{4}(\gamma-1)}{D_{1}^2}+\frac{d_{4}\gamma}{D_{2}^2}\right]+\nonumber\\&&
 \frac{(1+sw)\beta}{c}\left[\frac{6\mu(\gamma-1)+4(\gamma-1)^2 }{D_{1}^6} -\frac{1}{D_{1}^4}\right.\nonumber\\&&\left.-
 \frac{12\mu(\gamma-1)^3}{D_{1}^8}\right], \\\nonumber&&\\
 \nu_{191}&=&\frac{15l}{2}\left[\frac{c_{4}(\gamma-1)}{D_{1}^2}+\frac{d_{4}\gamma}{D_{2}^2}\right]+\nonumber\\&&
 \frac{(1+sw)\beta(1-\mu)}{c}\left[\frac{2\mu(\gamma-1)}{D_{1}^6}-\frac{1}{D_{1}^4}\right],\\\nonumber&&\\ 
\nu^{*}&=&1-mu-\gamma+\frac{c_{1}(\gamma-1)}{D_{1}}+\frac{d_{1}\gamma}{D_{2}}+\nonumber\\&&\frac{(1+sw)\beta(1-\mu)}{c}\left[
\frac{\mu l(1-\gamma)}{D_{1}^4}+\frac{l}{D_{1}^2}\right],
\end{eqnarray}
 \begin{eqnarray}
 \nu_{26}&=&-\frac{35l}{2}\left[\frac{c_{4}(\gamma-1)^3}{D_{1}^4}+\frac{d_{4}\gamma^3}{D_{2}^4}\right]+\nonumber\\&&
 \frac{15 l}{2}\left[\frac{c_{4}(\gamma-1)}{D_{1}^2}+\frac{d_{4}\gamma}{D_{2}^2}\right]+\frac{(1+sw)\beta (1-\mu)}{c}
 \times\nonumber\\&&\left[\frac{1}{D_{1}^4}- \frac{8(\gamma-1)^{2}+4\mu (\gamma-1)}{D_{1}^6} -\right.\nonumber\\&&\left.
 \frac{8(\gamma-1)^3 (1-\mu-\gamma)}{D_{1}^8}\right],\\\nonumber&&\\
 \nu_{27}&=&\frac{3}{2}(c_{4}+d_{4})+\frac{(1+sw)\beta(1-\mu)}{c}
 \frac{4l(\gamma-1-\mu)}{D_{1}^6},\nonumber\\&&\\
 \nu_{*}&=&c_2 +d_2-\frac{(1+sw)\beta(1-\mu)\mu l}{c D_{1}^4},\\\nonumber&&\\
 \nu_{29}&=&\frac{45l}{2}\left[\frac{c_{4}(\gamma-1)}{D_{1}^2}+\frac{d_{4}\gamma}{D_{2}^2}\right]+\nonumber\\&&
 \frac{(1+sw)\beta(1-\mu)}{c}\left[\frac{1}{D_{1}^4}-\frac{4(\gamma-1)^2}{D_{1}^6}\right],\\\nonumber&&\\
 \nu_{290}&=&\frac{15l}{2}\left[\frac{c_{4}(\gamma-1)}{D_{1}^2}+\frac{d_{4}\gamma}{D_{2}^2}\right]+\nonumber\\&&
 \frac{(1+sw)\beta(1-\mu)}{c}\left[\frac{1}{D_{1}^4}+\frac{4(\gamma-1)(1-\mu-\gamma)}{D_{1}^6}\right],\nonumber\\&&\\
 \nu_{291}&=&-\frac{15}{2}\left[\frac{c_{4}(\gamma-1)^2}{D_{1}^2}+\frac{d_{4}\gamma^2}{D_{2}^2}\right]+\frac{3}{2}
 (c_{4}+d_{4})\nonumber\\&&+
 \frac{(1+sw)\beta(1-\mu)}{c}\left[\frac{12l(\gamma-1)}{D_{1}^6}-\right.\nonumber\\&&\left.
 \frac{24l(\gamma-1)^{2}(\mu+1)}{D_{1}^{8}} \right],\\\nonumber&&\\
 \nu_{292}&=&\frac{3}{2}(c_{4}+d_{4})+\frac{(1+sw)\beta(1-\mu)}{c}\left[
 \frac{4l(\gamma-1)}{D_{1}^6}\right],\nonumber\\&&\\
 \nu^{**}&=&l\left[1-\left(\frac{c_{1}}{D_{1}}+\frac{d_{1}}{D_{2}}\right)\right]+\nonumber\\&&
 \frac{(1+sw)\beta(1-\mu)}{c}\frac{(\gamma-1+\mu)}{D_{1}^2},\\\nonumber&&\\
\nu_{32}&=&\frac{-15}{2}\left[\frac{c_{4}(\gamma-1)^2}{D_{1}^2}+\frac{d_{4}\gamma^2}{D_{2}^2}\right]+
 \frac{3}{2}(c_{4}+d_{4})+\nonumber\\&&
 \frac{(1+sw)\beta(1-\mu)\mu l}{cD_{1}^4}\left[\frac{12(\gamma-1)^2}{{D_{1}^4}}-\frac{2}{D_{1}^2}\right],\\\nonumber&&\\
 \nu_{33}&=&15l\left[\frac{c_{4}(\gamma-1)}{D_{1}^2}+\frac{d_{4}\gamma}{D_{2}^2}\right]+\nonumber\\&&
 \frac{4(1+sw)\beta(1-\mu)\mu (\gamma-1)}{cD_{1}^6},\\\nonumber&&\\
 \nu_{34}&=&\frac{3}{2}(c_{4}+d_{4})-\frac{6(1+sw)\beta(1-\mu)\mu l}{cD_{1}^6},\\\nonumber&&\\
 \nu_{35}&=&\frac{3}{2}(c_{4}+d_{4})-\frac{2(1+sw)\beta (1-\mu)\mu l}{c D_{1}^6},\\\nonumber&&\\
 \rho_{10}&=&-\frac{A_{14}}{\nu_{10}\nu_{21}-\nu_{20}\nu_{11}},
 \end{eqnarray}
 \begin{eqnarray}
 \rho_{11}&=&\frac{-A_{11}B_{11}+4\lambda A_{13}(\nu_{11}-\nu_{20})}{B_{11}^2 +16\lambda^2 (\nu_{11}-\nu_{20})^2},\\\nonumber&&\\
 \rho_{12}&=&\frac{B_{11}A_{12}}{B_{11}^2 +16\lambda^2 (\nu_{11}-\nu_{20})^2}\\\nonumber&&\\
 \rho_{13}&=&\frac{B_{11}A_{13}-4A_{12}\lambda(\nu_{11}-\nu_{20})}{B_{11}^2 +16\lambda^2 (\nu_{11}-\nu_{20})^2},\\\nonumber&&\\
 \rho_{14}&=&\frac{4A_{11}\lambda(\nu_{11}-\nu_{20})}{B_{11}^2 +16\lambda^2 (\nu_{11}-\nu_{20})^2},\\\nonumber&&\\
\rho_{20}&=&\frac{-\delta_{4}+\nu_{31}\rho_{10}}{\nu_{21}}, \\\nonumber&&\\
\rho_{21}&=&\frac{4\lambda\rho_{11}-\nu_{31}\rho_{14}-\delta_{1}}{-4\lambda^2-\nu_{21}},\\\nonumber&&\\
\rho_{22}&=&\frac{4\lambda\rho_{12}-\nu_{31}\rho_{13}}{-4\lambda^2-\nu_{21}},\\\nonumber&&\\
\rho_{23}&=&\frac{4\lambda\rho_{13}+\nu_{31}\rho_{12}+\delta_{3}}{-4\lambda^2-\nu_{21}},\\\nonumber&&\\
\rho_{24}&=&\frac{4\lambda+\nu_{31}\rho_{11}-\delta_{2}}{-4\lambda^2-\nu_{21}},\\\nonumber&&\\
\rho_{30}&=&\frac{h_{1}}{-3\lambda^2},\ \
\rho_{31}=\frac{h_{1}}{\lambda^2},\ \ \rho_{32}=\frac{h_{2}}{\lambda^2},\\\nonumber&&\\
\rho_{33}&=&\frac{h_{2}}{-3\lambda^2},\\\nonumber&&\\
A_{11}&=&4\lambda^2\alpha_{1} +\alpha{1}\nu_{21}+4\lambda\delta_{1}+\delta_{2}\nu_{11},\\\nonumber&&\\
A_{12}&=&4\lambda^2\alpha_{2}+\alpha_{2}\nu_{21}+\delta_{3}\nu_{11},\\\nonumber&&\\
A_{13}&=&4\lambda^2\alpha_{3}+\nu_{21}\alpha_{3}-4\delta_{2}\lambda+\delta_1\nu_{11},\\\nonumber&&\\
A_{14}&=&\alpha_{2}\nu_{21}+\nu_{11}\delta_{4},\\\nonumber&&\\
B_{11}&=&16\lambda^4-4\lambda^2(4-\nu_{21}+\nu_{10})+(\nu_{10}\nu_{21}-\nonumber\\&&\nu_{20}\nu_{11}),\\\nonumber&&\\
s_{11}&=&\alpha_{11}+2\omega_{2}\lambda^2 A_{x}(2\kappa-1),\\\nonumber&&\\
s_{12}&=&\alpha_{12}+2\kappa\lambda A_{x}\nu_{11}^2\omega_{2},\\\nonumber&&\\
\sigma_{10}&=&\frac{\beta_{11}\beta_{21}+\beta_{12}\beta_{22}}{\beta_{21}^2+\beta_{22}^2},\\\nonumber&&\\
 \sigma_{11}&=&\frac{\beta_{12}\beta_{21}-\beta_{11}\beta_{22}}{\beta_{21}^2+\beta_{22}^2},\\\nonumber&&\\
 \sigma_{12}&=&\frac{\beta_{13}\beta_{31}+\beta_{14}\beta_{32}}{\beta_{31}^{2}+\beta_{32}^2},\\\nonumber&&\\
 \sigma_{13}&=&\frac{\beta_{14}\beta_{31}-\beta_{13}\beta_{31}\beta_{32}}{\beta_{31}^2+\beta_{22}^2},
 \end{eqnarray}
 \begin{eqnarray}
c_{n}&=&\frac{(1-\beta)(1-\mu)}{D_{1}^{n+1}}, \ \ 
d_{n}=\frac{\mu}{D_{2}^{n+1}},\\\nonumber&&\\
 D_{1}&=&\sqrt{(\gamma-1)^2+l^2}, \ \
  D_{2}=\sqrt{\gamma^2 +l^2},\\ \nonumber&&\\ 
 \beta_{11}&=&-s_{11}\lambda^2 +\nu_{21}s_{11}+2s_{22}\lambda-\nu_{11}s_{21},\\\nonumber&&\\
 \beta_{12}&=&-s_{12}\lambda^2 -\nu_{21}s_{12}-2s_{21}\lambda-\nu_{11}s_{22},\\\nonumber&&\\
 \beta_{13}&=&-9\lambda^2 (\alpha_{13}+\zeta\alpha_{14})-\nu_{21}(\alpha_{13}+\zeta\alpha_{14})+\nonumber\\&&
 3\lambda(\alpha_{24}+\zeta\alpha_{26})-\nu_{11}(\alpha_{23}+\zeta\alpha_{24}),\\\nonumber&&\\
\beta_{14}&=&-9\lambda^2 (\alpha_{18}+\zeta\alpha_{16})-\nu_{21}(\alpha_{18}+\zeta\alpha_{16})-\nonumber\\&&
3\lambda(\alpha_{23}+\zeta\alpha_{24})-\nu_{11}(\alpha_{24}+\zeta\alpha_{26}),\\\nonumber&&\\
\beta_{21}&=&\lambda^4 -\lambda^2(4-\nu_{21}-\nu_{10})+\nu_{10}\nu_{21}+\nu_{11}\nu_{20},\\\nonumber&&\\
\beta_{22}&=&2\lambda(\nu_{11}+\nu_{20}),\ \ \beta_{32}=6\lambda(\nu_{11}+\nu_{20}),\\\nonumber&&\\
\beta_{31}&=&81\lambda^4 -9\lambda^2(4-\nu_{21}-\nu_{10})+\nu_{10}\nu_{21}+\nu_{11}\nu_{20},\\\nonumber&&\\
\sigma_{20}&=&\frac{-2\sigma_{11}\lambda+\nu_{20}\sigma_{10}+s_{21}}{-\lambda^2 -\nu_{21}},\\\nonumber&&\\
\sigma_{21}&=&\frac{2\sigma_{10}\lambda+\nu_{20}\sigma_{11}+s_{22}}{-\lambda^2 -\nu_{21}},\\\nonumber&&\\
\sigma_{22}&=&\frac{-6\sigma_{13}\lambda+\nu_{20}\sigma_{12}+\alpha_{23}+\zeta\alpha_{24}}{-9\lambda^2 -\nu_{21}},\\
\alpha_{1}&=&\frac{\nu_{21}A_{x}^2+\nu_{13}\kappa^2 A_{x}^2 \nu_{10}^2}{2}-2\nu_{13}\kappa^2 A_{x}^2 \lambda^2,\\\nonumber&&\\
\alpha_{2}&=&\frac{\nu_{14}A_{z}^2}{2},\ \  \alpha_{3}=2\nu_{13}\kappa^2 A_{x}^2 \lambda\nu_{10},\\
\alpha_{4}&=&\frac{\nu_{12}A_{x}^2 +\nu_{13}\kappa^2 A_{x}^2\nu_{10}^2}{2}+\frac{\nu_{14}A_{z}^2}{2}+
2\nu_{13}\kappa^2 A_{x}^2 \lambda^2,\\\nonumber&&\\
\delta_{1}&=&\nu_{25}\kappa A_{x}^2 \lambda-2\lambda\nu_{10}\nu_{23}\kappa^2A_{x}^2,\\\nonumber&&\\
\delta_{2}&=&\frac{\nu_{22}A_{x}^2-(\nu_{25}+\nu_{23}\kappa)\kappa A_{x}^2\nu_{10}}{2}-2\lambda^2\nu_{23}\kappa^2 A_{x}^2,\\
\delta_{3}&=&\frac{\nu_{24}A_{z}^2}{2},\\
\delta_{4}&=&\frac{(\nu_{22}-\nu_{25}\kappa \nu_{10}+\nu_{23}\kappa^2\nu_{10}^2 +2\lambda_{2}\kappa^2\nu_{23})A_{x}^2+
\nu_{24}A_{z}^2}{2},\nonumber\\&&\\
 h_{1}&=&\frac{(\nu_{30}+\nu_{31})A_{x}A_{z}}{2},\\\nonumber&&\\
h_{2}&=&\lambda\nu_{31}\kappa A_{x}A_{z}.
 \end{eqnarray}
 \label{lastpage}
\end{document}